\documentstyle[epsf,referee]{mn}
\begin{document}

\newcommand\approxgt{\mbox{$^{>}\hspace{-0.24cm}_{\sim}$}}
\newcommand\approxlt{\mbox{$^{<}\hspace{-0.24cm}_{\sim}$}}

\load{\scriptsize}{\sc}
\def\ion#1#2{\rm #1\,\sc #2}
\def\HI{{\ion{H}{i}}}
\def\HII{{\ion{H}{ii}}}
\def\GI{{\ion{He}{i}}}
\def\GII{{\ion{He}{ii}}}
\def\GIII{{\ion{He}{iii}}}
\def\MH{{{\rm H}_2}}
\def\Hp{{{\rm H}_2^+}}
\def\Hm{{{\rm H}^-}}

\def\dim#1{\mbox{\,#1}}

\def\figdir{.}

\def\Omeganow{{\Omega_0}}

\title[Equation of State of the IGM]{Equation of State of the Photoionized Intergalactic Medium}

\author[Hui and Gnedin]{Lam Hui and Nickolay Y.\ Gnedin\\
$^1$NASA/Fermilab Astrophysics Center, Fermi National Accelerator
Laboratory, Batavia, IL 60510\\
$^2$Department of Astronomy, University of California, Berkeley, 
CA 94720} 

\maketitle

\begin{abstract}
We develop an efficient method to study the effects of reionization history on
the temperature-density relation of the intergalactic medium in 
the low density limit
(overdensity $\delta \, \approxlt \, 5$). It is applied to the study of
photo-reionization models in 
which the amplitude, spectrum and onset epoch of the ionizing flux, as well 
as the cosmology, are systematically varied. We find that the mean
temperature-density relation at $z = 2-4$ is well approximated by a power-law
equation of state for uniform reionization models.  We derive analytical
expressions for its evolution and exhibit its asymptotic behavior: it
is found that for sufficiently early reionization, imprints of reionization
history prior to $z \sim 10$ on the temperature-density relation are
washed out. In this limit the temperature at cosmic mean density is
proportional to 
$[{\Omega_b h/\sqrt\Omeganow}]^{1/1.7}$. While the amplitude of the radiation
flux at the ionizing frequency of $\HI$ is found to have a negligible effect on the temperature-density relation 
as long as the universe reionizes before $z \sim 5$, the spectrum can change 
the overall temperature by about $20 \%$, through variations in the abundances of helium species. However the slope of the mean equation of  
state is found to lie within a narrow range for all reionization models we study,
where reionization takes place before $z \sim 5$. We discuss the implications
of these findings for the observational properties of the Ly$\alpha$
forest. In particular, uncertainties in the temperature of the intergalactic
medium, due to the uncertain reionization history of our universe, introduces a
$30 \%$ scaling in the amplitude of the column density distribution while the 
the slope of the distribution is only affected by about $5 \%$. 
Finally, we discuss how a fluctuating ionizing field affects the above results.
We argue that under certain conditions, the loss of memory of reionization
history implies that at late times, the temperature-density relation of a gas
in a fluctuating ionizing background can be approximated by one that results
from a uniform radiation field, provided the universe 
reionizes sufficiently early. 
\end{abstract}


\def\tablefit{
\begin{table}
\label{tabfit}
\caption{Photoionization cross-section parameters}
\medskip
\[
\begin{tabular}{cccccccc}
Species & $E_0$ ($\dim{eV}$) & $\sigma_0$ ($\dim{cm}^2$) & 
$P$ & $y_a$ & $y_w$ & $y_0$ & $y_1$ \\
\noalign{\hrule}
$\HI$  & $4.298\times10^{-1}$ & $5.475\times10^{-14}$ & $2.963$ & $32.88$ & $0    $ & $0     $ & $0    $ \\
$\GI$  & $1.361\times10^{ 1}$ & $9.492\times10^{-16}$ & $3.188$ & $1.469$ & $2.039$ & $0.4434$ & $2.136$ \\
$\GII$ & $1.720             $ & $1.369\times10^{-14}$ & $2.963$ & $32.88$ & $0    $ & $0     $ & $0    $ \\
\end{tabular}
\]
\end{table}
}

\section{Introduction}

Absorption line studies of the low column density Ly$\alpha$ forest ($N_\HI
\, \approxlt \, 15 \dim{cm}^{-2}$; 
see Hu et al.\ 1995; Lu et al.\ 1996; Cristiani et al.\
1996) offer a unique probe of the universe which is free of luminosity
bias, spans a wide redshift range $z \sim 2-4$ and where the overdensity is
mildly nonlinear (overdensity $\delta \, \approxlt \, 5$, see Hui, Gnedin \&
Zhang 1996) and hence the relevant physics is simple to understand. Recent
hydrodynamic simulations support the hypothesis that the Ly$\alpha$ forest
arises from a fluctuating intergalactic medium which is the natural result of
gravitational instability in hierarchical clustering cosmological models
(Cen et al.\ 1994; Zhang et al.\ 1995; Hernquist et al.\ 1996;
Miralda-Escud\'{e} et al.\ 1996; Wadsley \& Bond 1996). 

The main quantity of interest in such studies is the Ly$\alpha$ transmission
(or optical depth) as a function of frequency, which is determined by the
density of neutral hydrogen as a function of position. 
The Gunn-Peterson effect (Gunn \& Peterson
1965) tells us that hydrogen (as well as helium) is highly ionized for $z \,
\approxlt \, 5$. Several models for its origin have been studied:
photoionization by first generation of stars or 
quasars, collisional ionization induced by shock-heated gas, decaying neutrinos
and so on (Ikeuchi \& Ostriker 1986; Couchman \& Rees 1986; Sciama 1990;
Miralda-Escud\'{e} \& Ostriker 1992; Shapiro, Giroux, \& Babul 1994; 
Giroux \& Shapiro 1996; Haiman \& Loeb
1996). The reader is also referred to Shapiro (1995) for an excellent review on
the subject. We focus on photoionization models in this
paper although the semi-analytical method we propose in this paper can be
applied to any model. 

Ionization equilibrium, which should hold after the universe reionizes,
dictates that the neutral hydrogen density is approximately proportional to
$\rho_b^2 T^{-0.7}$ where $\rho_b$ is the baryon density and $T$ is the
temperature (see 
\S~\ref{analytical}). It is therefore important 
to know the amplitude of $T$ and
its dependence on $\rho_b$. As we will discuss later, the temperature-density
relation can significantly affect observed properties of the Ly$\alpha$
forest, its column density distribution for instance. 

Imagine a fluid element evolving in a photoionized intergalactic medium.
Its temperature at $z\sim 2-4$ is influenced by a number of factors: how its
density evolves with time, which determines the amount of
heating/cooling; how and when the ionizing radiation is turned on; the
radiation amplitude and 
spectrum, which control how much photoionization heating the fluid element
suffers; cosmological parameters such as baryon density, the Hubble constant
and the density parameter $\Omega_0$ which affects the recombination rate,
adiabatic cooling rate and the evolution of density. 

The temperature-density relation for a set of fluid elements is then
determined by, among other things, the reionization history of the
universe. Given a reionization 
scenario and cosmological model, hydrodynamic simulations can predict this
relation accurately, but 
limited computer resources obviously restrict the number of reionization
histories one can study. It is therefore 
important that we develop alternative tools in order to understand the effects
of a full
range of reionization histories, considering the as yet poor knowledge of the
actual 
reionization history of our universe. Moreover, it is worth pointing out that
there exists analytical tools borrowed from studies of large scale structure
which can predict the distribution of $\rho_b$ with reasonable accuracy in the
low
density regime (Bi,
Borner \& Chu 1992; Bi \& Davidsen 1996; Gnedin \& Hui 1996; Hui, Gnedin \&
Zhang 1996). By their analytical/semi-analytical nature, it is possible to
study a whole range of cosmological models in an efficient manner. What remains
to be specified  
for these methods, in order to calculate quantities like the column density
distribution of hydrogen, is the temperature-density relation. A calculation
aiming at just that is attempted here.

We propose to model the density evolution of each fluid element using the
Zel'dovich approximation (Zel'dovich 1970), which is known to be a good
approximation in the mildly nonlinear regime (Coles, Melott, \& Shandarin
1993) and for elements where
hydrodynamic effects like pressure and shock-heating are unimportant.
Given a probability distribution of the initial configuration for these
elements, which can be deduced for any given cosmological model,
one can generate a whole set of such elements and simply follow their
thermal and chemical evolution. The resulting bulk temperature-density relation
can then be studied for any chosen reionization history. The virtue of this
method is that each element can be treated independently and that 
the density evolution is given by an analytical formula (from the Zel'dovich
approximation) while the thermal
and chemical evolution equations involve a set of ordinary differential
equations which can be solved numerically. Obviously, such a Lagrangian method
does not incorporate the effects of shock-heating properly. We check 
the results of our method against full hydrodynamic simulations and find
good agreement for $\delta \approxlt 5$, implying 
shock effects are not significant in the low density regime (\S~\ref{simulation}). 

Most previous work investigating the effects of different reionization
histories either assumes a uniform intergalactic medium (eg.\ Giroux \& Shapiro
1996) or uses the spherical collapse model for density evolution
(eg.\ Miralda-Escud\'{e} \& Rees 1994). For the low column density Ly$\alpha$ forest,
which arises naturally from a mildly fluctuating intergalactic medium, and
which hydrodynamic simulations indicate to be consisting mostly of filaments
and pancakes (Cen et al.\ 1994), the density evolution is better approximated
by the Zel'dovich approximation (Hui et al.\ 1996).

A great deal of effort has been made to examine radiative
transfer effects after the onset of ionizing radiation, including the
expansion of $\HII$ regions, reprocessing by absorption systems etc.\ (Zuo
1992a,b; Meiksin \&
Madau 1993; Miralda-Escud\'{e} \& Rees 1994; Giroux \& Shapiro 1996; Haardt \&
Madau 1996; see also Gnedin \& Ostriker 1996 for numerical 
simulations that include some of these effects). These effects can
in principle be incorporated in our method 
by specifying how the radiation intensity correlates with density and possibly
allowing the onset of reionization to occur at different times for different
fluid elements. While more work still needs to be done in order to make
such a specification with reasonable accuracy, we focus in this paper on
reionization models in 
which the ionizing flux is uniform, but allowing the onset-epoch, spectrum and
amplitude of the ionizing flux as well as the cosmology to vary. It turns out
one can learn something about the more realistic fluctuating case, even from
these simple models, which we will discuss in \S~\ref{discuss}. 

As we will show, the mean temperature-density relation
of the photoionized intergalactic medium in the low density regime is
well-approximated by a power law equation of state ($T \propto
[1+\delta]^{\gamma-1}$) where $\delta$ is the 
mass overdensity). The questions we would like to address are then: how
does the amplitude and slope of the equation of state depend on the
reionization history, characteristics of the ionizing radiation as well as
cosmological parameters; what are the
implications for observational properties of the Ly$\alpha$ forest?

Organization of the paper is as follows. 
In the next section, we outline a semi-analytical method to predict  the 
temperature-density relation for any given
cosmology and reionization history, using the Zel'dovich approximation. 
An example is discussed and a comparison is made with the result of a full 
hydrodynamic simulation, computed using the same reionization history.
Then, in \S~\ref{sudden} we apply the method to study models in
which the universe reionizes suddenly, due to a rapid outburst of radiation. We
investigate systematically how 
varying the epoch of reionization, cosmological parameters and the radiation
amplitude and 
spectrum affects the equation of state at later redshifts. 
Analytical approximations to the equation of state are derived.
In \S~\ref{preheat}, it is shown that models in which reionization
occurs after a period of reheating (gradual turn-on of radiation) give
equations of state that can be  
well-fitted by that of an appropriate sudden reionization model.
Finally we discuss the implications of our findings for the observational
properties of the Lyman-alpha Forest and the intergalactic medium in
\S~\ref{discuss}. A summary of relevant reaction rates is provided in the 
Appendix.

A few words about our terminology. The three different methods or procedures
to study the effects of reionization history, are respectively referred to 
as ``analytical approximation'', ``semi-analytical method'' and ``hydrodynamic
simulation''. The first is discussed in \S~\ref{sudden}, where a few
approximations are made to derive the equation of state in closed form;
the second is explained in \S~\ref{equation}, which evolves the density
analytically but requires numerical integration of the thermal and chemical
evolution equations, though not a full scale hydrodynamic 
simulation and the third is what its name suggests. While the last method
is perhaps the most accurate (provided the numerical resolution is
sufficient), the second, which is the focus of this paper,
allows us to study the temperature-density relation for a large number
of reionization histories in an efficient and accurate manner, albeit
restricted 
to the low density regime; the first allows us to give useful approximate
quantitative expressions that embody the correct qualitative trends.
Standard symbols are used for cosmological
parameters: $a = (1+z)^{-1}$ where $1+z$ is the cosmological redshift factor,
$H$ for the Hubble constant as a function of $z$, $H_0$ 
for the Hubble constant today, $h$ for $H_0 / 100 \dim{km}\dim{s}^{-1}
\dim{Mpc}^{-1}$, $\Omega$ for the density parameter today, with the
subscript $b$ to denote its baryon portion and $0$ all its matter
content, and $\Lambda$ for the contribution from cosmological
constant. We use the symbol ${\rm h}$ (in distinct from $h$) to denote the
Planck constant in a few places where it arises and $k_{\rm B}$ to denote
the Boltzman constant.

\section{The Semi-analytical Method}
\label{method}

\subsection{The Equations}
\label{equation}

In a nutshell, the method we propose is very simple: assume that the density
evolution of a given fluid element is governed by the Zel'dovich 
approximation; allow the element to cool/heat and its constituents to 
change according to standard thermal and chemical reaction rates; 
repeat the same procedure for a whole set of fluid elements with
initial parameters drawn from a distribution implied by whatever
cosmological initial conditions one prefers (in this paper, we study only 
Gaussian random initial conditions). The resulting temperature-density
relation can be obtained by 
making a scatter-plot of temperature versus density for all the fluid
elements at a given time of interest.

The density evolution of a fluid element, according to the 
Zel'dovich approximation, is given by:
\begin{equation}
1 + \delta = \,{\rm det}^{-1} \left[\delta_{ij} + D_+(t)\psi_{ij} \right] \, ,
\label{zadelta}
\end{equation}
where $\delta$ is the mass overdensity, $\delta_{ij}$ is the Kronecker delta 
and $D_+(t)$ is the linear growth factor (Peebles 1980), the time dependence
of which is completely specified by cosmology (e.g. it equals the Hubble
scale factor $a$ for a universe at critical matter density). The normalization 
is chosen such that $D_+ = 1$ today.

The tensor $\psi_{ij}$ is a $3 \times 3$ symmetric matrix determined by 
initial
conditions (i.e. its components vary from one fluid element to another). One
can  
always choose a basis in which the matrix is diagonal. So, the problem of 
initial conditions reduces to finding the probability distribution of its 
three eigenvalues. This problem was solved by Doroshkevich (1970) for
Gaussian random initial density field, (see also Bartelmann and 
Schneider (1992) for an elegant alternative derivation) which we also assume
in this paper. For recent discussions
of how to generate realizations of matrix $\psi_{ij}$, see Bertschinger and
Jain (1994) and Reisenegger and Miralda-Escud\'{e} (1995). It can be shown
that the probability distribution of the matrix $\psi_{ij}$ is uniquely
specified by the dispersion of its trace. In the limit where $D_+(t)$ is 
small, it can be seen that the trace multiplied by $- D_+(t)$ is equal to 
$\delta$ i.e. the linear overdensity. Therefore, the average of the trace 
vanishes and its rms (root mean squared) value is equal to the rms linear
density fluctuation today $\sigma$. 
At a given redshift and for a fixed number of fluid elements, a given rms $\sigma$
value would imply a probable range of $\delta$'s in which most elements
will end up. For the range we are interested in ($\delta$ between about 
$-0.9$ and $5$, at $z$ from $2$ to $4$), we find that setting the rms value
$\sigma$ to $2$ is adequate. For a fixed number of fluid elements, choosing a
higher value for $\sigma$ has the effect of shifting 
the probable range of densities upward and also increasing
the scatter in a plot of temperature versus density. Alternatively,
for a fixed rms value $\sigma$, one can reproduce 
this latter effect and increase the number of
fluid elements that fall in the desired density range by increasing the
total number of fluid elements in a realization. We find that for our densities
and redshifts of 
interest, using a different $\sigma$ does not change the overall
temperature-density relation significantly. But it should be kept in mind the
scatter in temperature-density relation is limited, in part, by the finite number of
elements in the realizations.

Temperature obeys the following equation of evolution:
\begin{equation}
{dT\over dt} = -2HT + {2T\over{3(1+\delta)}} {d\delta\over dt} -{T\over \sum_i
\tilde X_i} {d\sum_i \tilde X_i\over dt} + {2\over {3k_B n_b}}
{dQ\over dt} \, ,
\label{T}
\end{equation}
where $d/dt$ is the Lagrangian derivative following each fluid
element, $n_b$ is the proper number density of all gas particles
(i.e. everything except noninteracting dark matter) and $T$ is the
temperature which depends on both space and time. The symbol $\tilde X_i$ is
defined by $n_i \equiv (1+\delta) \tilde X_i\, {\bar\rho_b/m_p}$,
where $n_i$ is the proper number density of the specie $i$, $\bar\rho_b$ 
is the mean mass density of baryons at the time of interest and $m_p$
is the mass of the proton and $\delta$ is the mass overdensity. 
For example the neutral 
fraction of hydrogen, $X_\HI$ (to be distinguished from $\tilde
X_\HI$), is then $\tilde X_\HI/(\tilde X_\HI+\tilde X_\HII)$. 
Note that $\tilde X_i$ is a function of space and time in general.   

The first two terms on the right hand side take care of adiabatic cooling
or heating. The third accounts for the change of internal energy per particle
due to the change in the number of particles. The last term $dQ/dt$ is the
heat gain (or negative hear loss) per unit volume by the gas particles
from the surrounding radiation field. The heating and cooling rates
due to photoionization, recombination and compton scattering are summarized
in the Appendix.

Equation (\ref{T}) has to be supplemented by one that determines the
abundance of each particle type, which takes the form:
\begin{equation}
{d\tilde X_i \over dt} = -\tilde X_i \Gamma_i + \sum_{j,k} \tilde X_j \tilde
X_k 
R_{jk} \left[{\bar\rho_b (1+\delta)\over m_p}\right]
\, ,
\label{n}
\end{equation}
where $\Gamma_i$ is the photoionization rate of the specie $i$ and $R_{jk}$ is
the recombination rate (in units of volume per unit time) of the species
$j$ and $k$ to give $i$. The photoionization rate is given by
\begin{equation}
\Gamma_i = \int_{\nu_{i}}^{\infty} 4 \pi J_\nu \sigma_{i} {d\nu\over
{\rm h}\nu}  \, ,
\label{P}
\end{equation}
where $J_\nu$ is the specific intensity of the ionizing radiation as
a function of frequency $\nu$, $\nu_i$ is the frequency above which
a photon can ionize the specie $i$ and $\sigma_i$ is the cross-section
for this process. The cross-sections for different species as well as
the recombination rates are given in the Appendix.

What remains to be specified is the time evolution of the specific intensity
of the ionizing radiation as a function of frequency, which
determines the photoionization ($\Gamma_i$) and photoionization heating
($dQ/dt$) rates. We will examine the effects of a wide range of amplitudes,
spectra and time evolution of the ionizing radiation
in this paper. 

Equations (\ref{zadelta}), (\ref{T}), and (\ref{n}) completely determine the
thermal  
and chemical evolution of a fluid element. They are numerically integrated
for each fluid element with the initial condition for $\delta$ determined
as discussed before. The thermal and chemical initial conditions 
are chosen as follows. The gas temperature $T$ is
equal to the cosmic microwave background temperature at $1+z = 100 
(\Omega_b h^2
/0.0125)^{2/5}$ (maintained by Compton scattering; Peebles 1993) 
and evolves adiabatically
after that until the universe is reionized by the UV
background. Abundances are assumed to be primordial, which is consistent
with observations so far in the low density intergalactic medium.
All species are neutral to high accuracy until reionization occurs. One can 
integrate equations (\ref{T}) and (\ref{n}) forward starting
from any time between $z = 100$ and the beginning of reionization.

\subsection{Comparison with Hydrodynamic Simulation}
\label{simulation}

\def\capAab{
A a scatter-plot (black dots) of temperature
versus density 
of 40000 cells randomly drawn from a $64^3$ CDM+$\Lambda$ hydrodynamic
at $z = 4$ using the SLH-P$^3$M  algorithm (see Gnedin 1995
and 
Gnedin \& Bertschinger 1996) with 
$\Omeganow = 0.35$, $\Omega_\Lambda = 0.65$, 
$\Omega_b = 0.055$ and $h = 0.7$ ({\it left panel\/}). 
The evolution of the ionizing background for the simulation is shown in
shown in Fig.\ \protect{\ref{fig2ab}}. 
The right panel shows the
temperature-density 
relation obtained for exactly the same  
cosmological and radiation parameters using the semi-analytical method 
described in \S~\protect{\ref{equation}}, applied to 2000 fluid elements. 
The larger scatter in Fig.\ \protect{\ref{fig1ab}}a compared to 
\protect{\ref{fig1ab}}b is in part due to the smaller
number of elements in the latter. The solid line plotted in  
both figures correspond to the analytical approximation expressed in
equations (\protect{\ref{Texpand}}), 
(\protect{\ref{T1}}) 
and (\protect{\ref{T0modified}}) for a sudden reionization model where
the epoch of reionization is $z = 9.6$.
}
\begin{figure}
\par\centerline{%
\epsfxsize=0.5\columnwidth\epsfbox{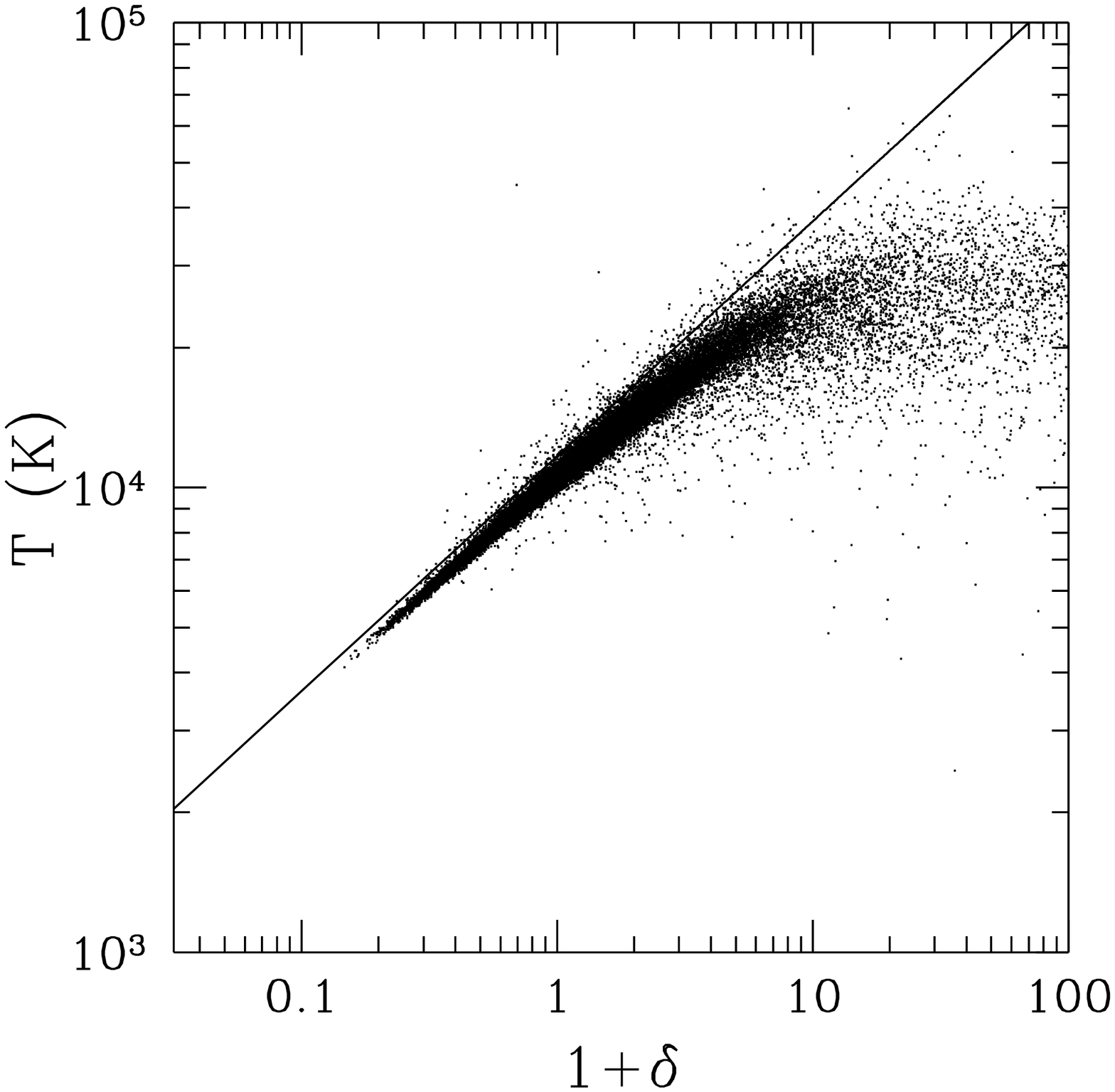}\hfil%
\epsfxsize=0.5\columnwidth\epsfbox{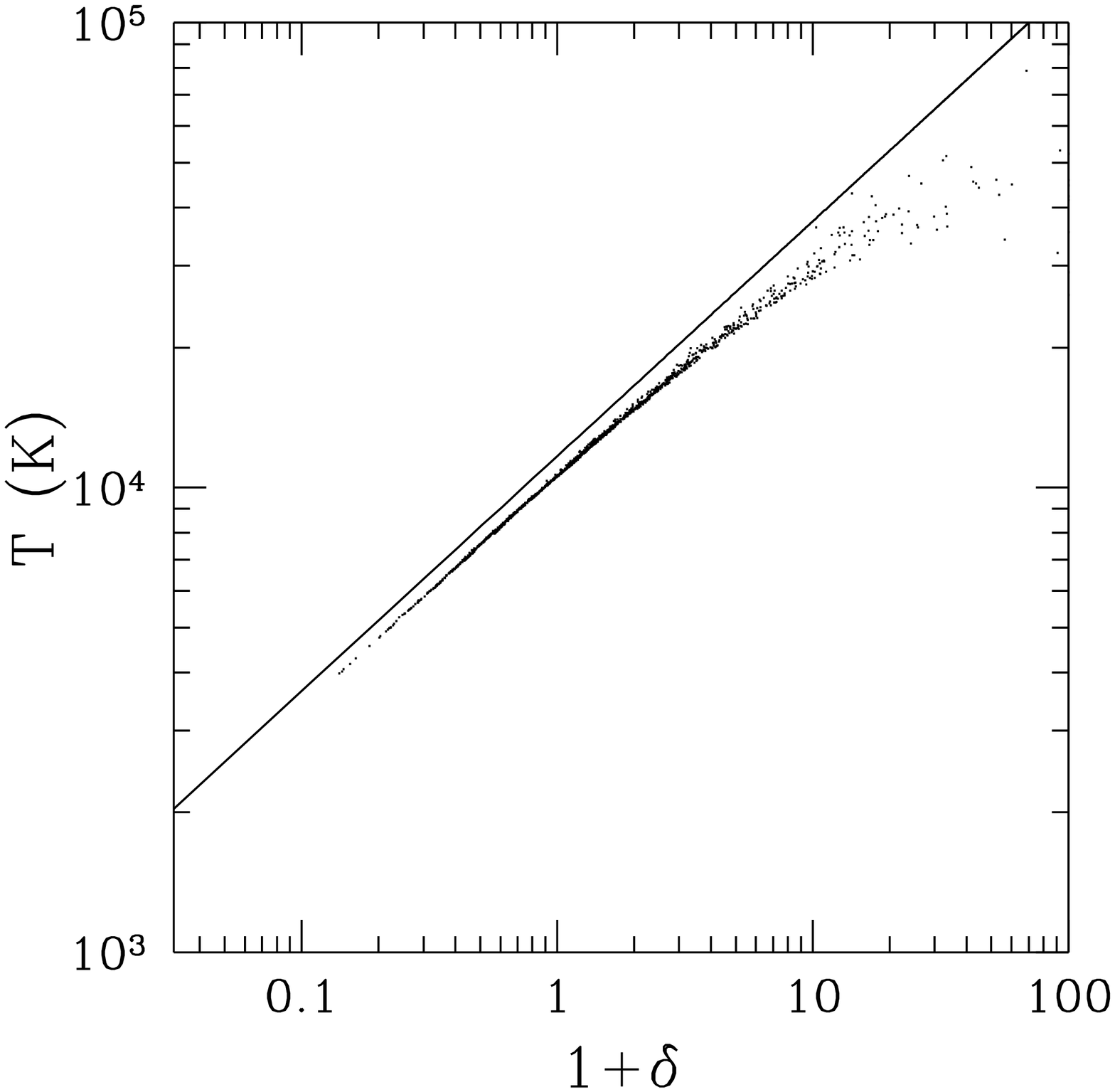}}%
\caption{\capAab}\label{fig1ab}
\end{figure}

\def\capBab{
Evolution of $J_\HI$ ({\it solid line\/}), $J_\GI$ ({\it 
dashed line\/}) and $J_\GII$ ({\it dotted line}) as a function of 
redshift, used in the computations shown in Fig.\ \protect{\ref{fig1ab}}.
The $J_i$'s are defined in equation (\protect{\ref{JHI}}).
}
\begin{figure}
\par\centerline{%
\epsfxsize=0.75\columnwidth\epsfbox{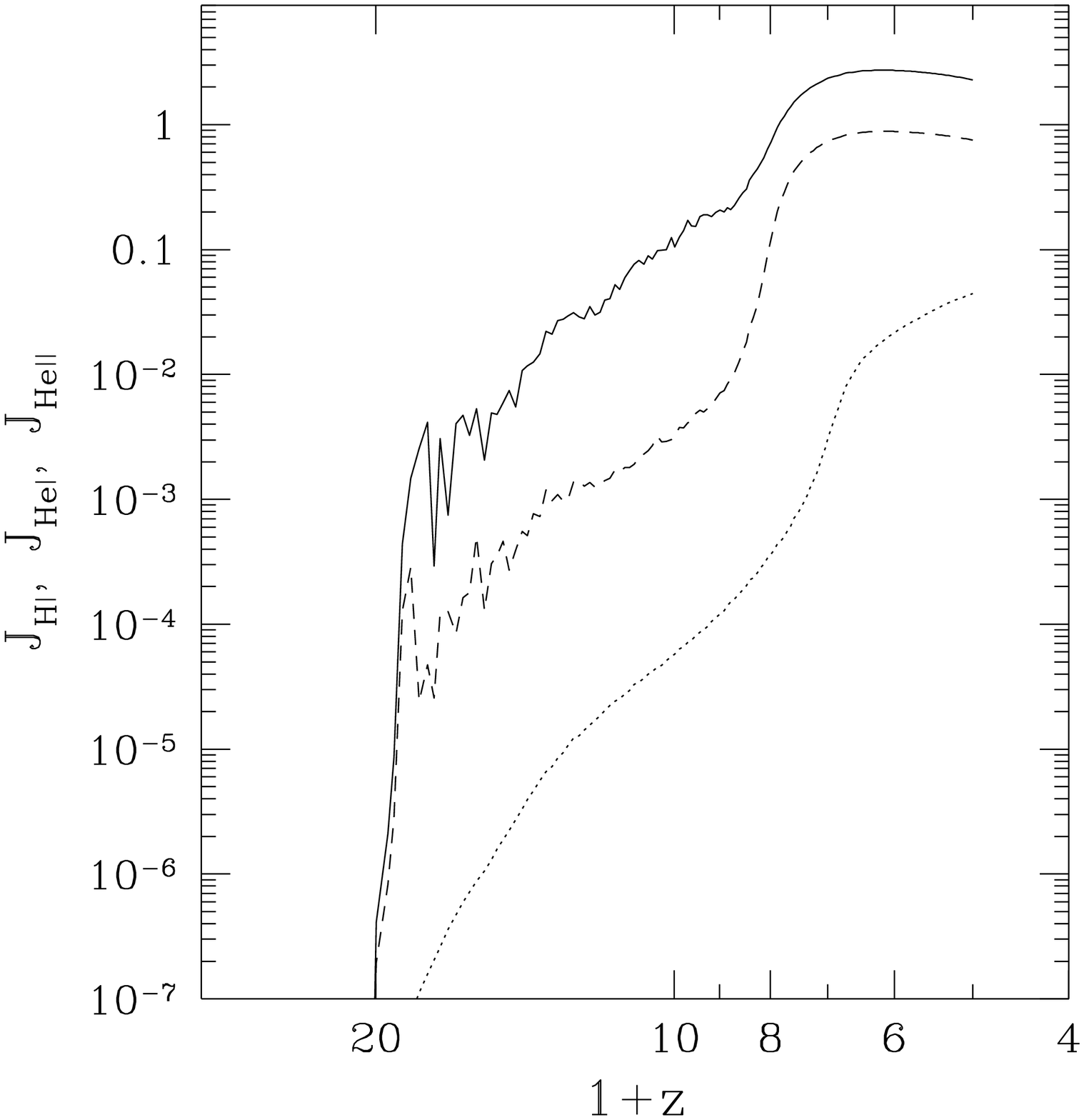}}%
\caption{\capBab}
\label{fig2ab}
\end{figure}

We show in Fig.\ \ref{fig1ab} scatter plots of temperature versus density
at $z = 4$ obtained using the semi-analytical method outlined above and 
using a full hydrodynamic simulation respectively. The same uniform ionization
field  has been used for both of them, the evolution of which is shown in
Fig.\ 
\ref{fig2ab}.\footnote{See \S~\ref{discuss} for a discussion of the effect 
of a non-uniform radiation field.}  The radiation field is computed in a
hydrodynamic 
simulation that includes the effects of star formation (see Gnedin and 
Ostriker 1996). In Fig.\ \ref{fig2ab}2, 
we quantify the level of the ionizing radiation by
the photoionization rates it implies i.e.
\begin{equation}
J_{i} = {{\int_{\nu_{i}}^{\infty} 4 \pi J_\nu \sigma_{i} {d\nu/\nu}  
\over {\int_{\nu_{i}}^\infty 4 \pi \sigma_{i} {d\nu/\nu}}}
(10^{-21} \dim{ergs}\dim{Hz}^{-1}\dim{s}^{-1}\dim{cm}^{-2}\dim{ster}^{-1}
)^{-1}}\, ,
\label{JHI}
\end{equation}
where the notations are the same as those defined in equation (\ref{P}).
For instance, $J_\HI$ is related to the photoionization rate of neutral
hydrogen $\Gamma_\HI$ by $\Gamma_\HI \sim  4 \times
10^{-12} J_\HI\dim{s}^{-1}$ (A more accurate proportionality constant 
can be inferred from the cross-section $\sigma_\HI$ given in our
Appendix). The three different quantities $J_\HI$, $J_\GI$ and
$J_\GII$ shown in Fig.\ \ref{fig2ab} do not completely specify $J_\nu$,
but they give an indication of how $J_\nu$ changes with time at different
frequencies. The full spectrum of $J_\nu$ is of course known and used in
producing the results shown in Fig.\ \ref{fig1ab} but we do not show it here.
(In practice, one actually needs only three more pieces of information
regarding $J_\nu$  for the thermal evolution of a gas of primordial  
composition: the photoionization {\it heating} rates of $\HI$, $\GI$
and $\GII$.)

Our simple semi-analytical method gives a temperature-density relation which
is well-approximated by a power law for $\delta
\,\approxlt\, 5$. It gives the correct mean behavior of gas elements in the
full hydrodynamic simulation in the low density regime. The larger scatter in
Fig.\ \ref{fig1ab}a 
compared to 1b is in part due to the smaller number of fluid 
elements in the latter. At higher densities, one begins to see the effect of
shocks: a wide scatter of temperature at a fixed density. The calculation 
shown in  Fig.\ \ref{fig1ab}b 
is not reliable in this regime. This is because by 
assuming that density evolves as prescribed by the Zel'dovich approximation, 
which is only a good approximation for pressureless (dark) matter or 
baryons at large scales, one misses the effects of gas pressure and shocks
as the fluid element is compressed to sufficiently high densities.

We have made a number of similar comparisons between the temperature-density
relations obtained from hydrodynamic simulations versus using our
semi-analytical method, for a number of different $J_\nu$'s as a function of
time. The semi-analytical method consistently gives the correct mean
behavior of gas elements of low density. Keeping in mind the intrinsic
scatter such as that seen in Fig.\ \ref{fig1ab}a, 
we can make use of our simple
semi-analytical method to efficiently study the mean temperature-density
relation at the low density regime for a large number of reionization
scenarios, which is the subject of the next two sections. 

\section{Sudden Reionization Models}
\label{sudden}

\subsection{Variation with the Epoch of Reionization}
\label{epoch}

\def\capCabcd{
The temperature-density relation for 4 different
sudden-reionization 
models: sudden reionization (see eq. [\protect{\ref{Jevol}}])    
at $z = 5$ ({\it a\/}), $z = 7$ ({\it b\/}), $z =
10$ ({\it c\/}), and $z = 19$ ({\it d\/}). For each reionization model, the black
dots shown are the results of calculations using the semi-analytical method
outlined in 
\S\protect{\ref{equation}} for $2000$ elements, shown at three different
instants: $z$ being 4, 3 and 2 from top to bottom.
The cosmological parameters are $h = 0.5$, $\Omeganow = 1$ and
$\Omega_b h^2 = 0.0125$ (with primordial abundances for hydrogen and helium).
The ionizing background is specified by its amplitude $J_\HI = 0.5$
(eq. [\protect{\ref{JHI}}]) and spectrum obeying equation 
(\protect\ref{cutoff})
with $f = 0.01$. The solid, dotted, and  dashed lines, from top to
bottom for 
each reionization 
model, represent the analytical expressions for the equation of state in
equations (\protect{\ref{Texpand}}), (\protect{\ref{T1}}), and
(\protect{\ref{T0modified}}), using the corresponding cosmological parameters
and reionization-epoch for each model. The only exception is panel ({\it d\/})
 where the
lines shown are exactly the same as those in the panel ({\it c\/}), i.e.\ 
setting $a_{\rm
reion} = 1/11$ in equations (\protect{\ref{T1}}) and 
(\protect{\ref{T0modified}})
(see explanation in the text).
}
\begin{figure}
\par\centerline{%
\epsfxsize=0.90\columnwidth\epsfbox{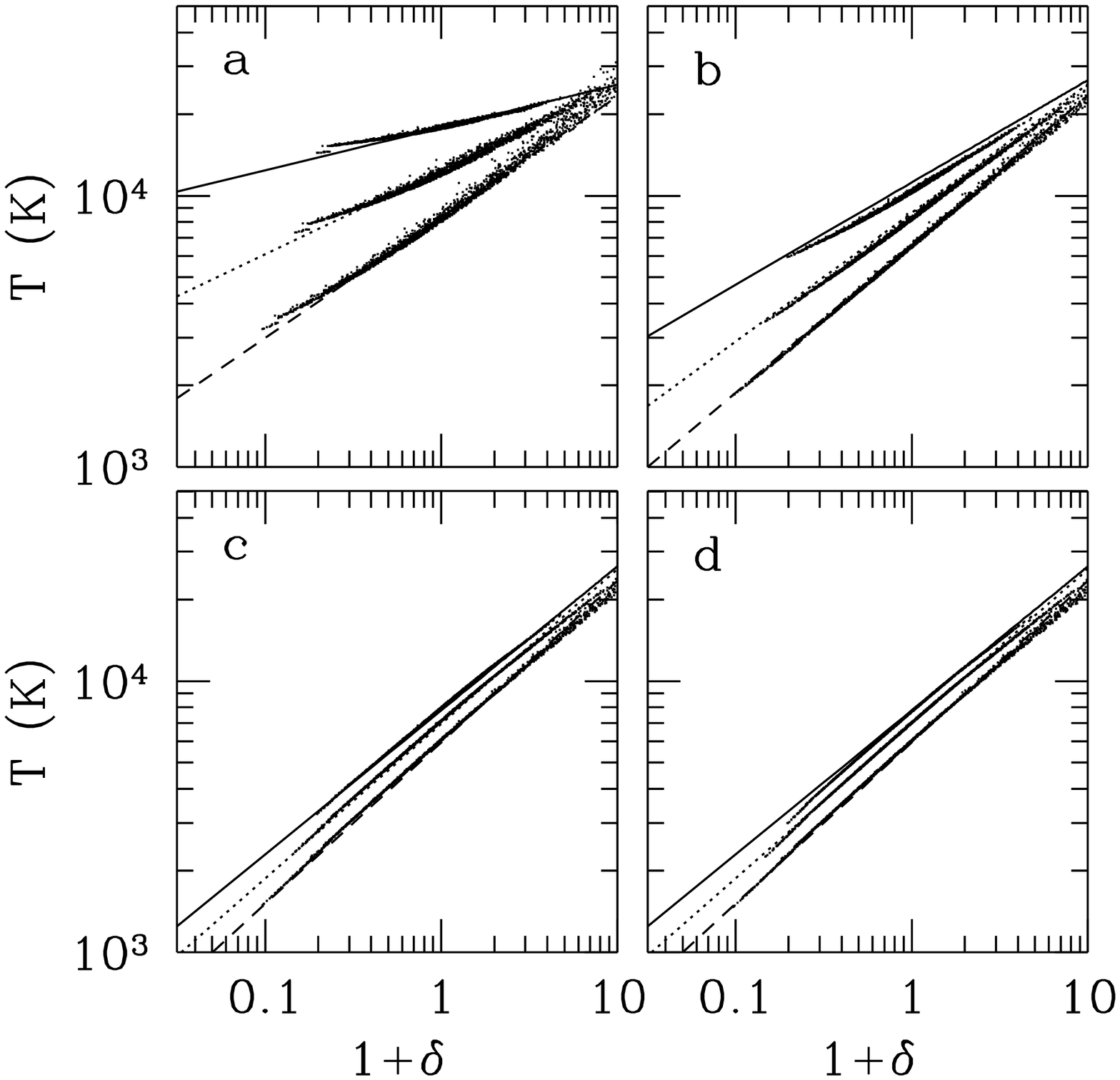}}%
\caption{\capCabcd}
\label{fig3abcd}
\end{figure}

In Fig.\ \ref{fig3abcd}a,b,c and d, 
we show the temperature-density relation for
four different sudden reionization models, where the reionization epoch
is systematically varied. All of them have no ionizing background until
the specified epoch and then $J_\HI$ (eq. [\ref{JHI}]) is taken to 
be $0.5$ from thereon i.e.
\begin{equation}
J_\HI = \cases{J_{\rm ion} & for $a\geq a_{\rm reion}$, and\cr
0 & for $a < a_{\rm reion}$,}
\label{Jevol}
\end{equation}
where $a_{\rm reion} = (1+z_{\rm reion})^{-1}$ is the Hubble scale factor when
the radiation is turned 
on and the parameter $J_{\rm ion}$ is chosen to be $0.5$ (we will study
the effects of varying $J_{\rm ion}$ later on).

It is assumed that $J_{\nu} \propto \nu^{-1}$, except
that it suffers a factor of $f = 0.01$ reduction beyond the frequency
corresponding to the ionization of $\GII$:
\begin{equation}
J_\nu = \cases{\phantom{f}J_0 \left({\nu\over\nu_\HI}\right)^{-1} &
for $\nu  \leq \nu_\GII$, and\cr
f J_0 \left({\nu\over\nu_\HI}\right)^{-1} & for $\nu > \nu_\GI$,}
\label{cutoff}
\end{equation}
where $J_0$ is chosen so that $J_\HI$ (eq. [\ref{JHI}]) has the desired
value, $\nu_\GII$ is the frequency corresponding to the ionization of
$\GII$ and $f$ (now chosen to be $0.01$) is a parameter specifying by 
how much the ionizing radiation is to be further diminished beyond 
$\nu_\GII$. 

Ignoring for the time being the solid, dotted and dashed lines in
Fig. \ref{fig3abcd} which will be discussed in the next section, the first
trend to notice is that the earlier the epoch of reionization, the 
lower the overall temperature at a given 
redshift, except that for sufficiently early reionization ($z_{\rm reion} >
10$) , the temperature-density relation approaches an asymptote. 
The second, somewhat striking, result is how well the temperature-density
relations are approximated by power-law equations of state, where the slope
steepens as universe reionizes earlier. The third is that when the 
reionization
epoch is pushed sufficiently early, the exact time when it occurs makes
very little difference to the equation of state: hence the very similar
temperature-density relations for reionization starting at $z = 10$ and $z =
19$ in Fig.\ \ref{fig3abcd}c and \ref{fig3abcd}d. 

To understand these trends, we perform the following approximate analytical
calculation.

\subsubsection{Analytical Calculation}
\label{analytical}

First, let us determine the reionization temperature. 
In the limit when a substantial amount of radiation is suddenly turned on, one
can approximate equations (\ref{T}) and (\ref{n}) by:
\begin{eqnarray}
{dT\over dt} &=& -{T\over \sum_i
\tilde X_i} {d\sum_i \tilde X_i\over dt} + {2\over {3k_B}} {\tilde X_{\rm
HI}\over \sum_i \tilde X_{i}} 
\int_{\nu_\HI}^{\infty} 4 
\pi J_\nu \sigma_\HI ({\rm h} \nu - {\rm h} \nu_\HI) {d\nu\over {\rm
h}\nu} \\ \nonumber {d \tilde X_\HI \over dt} &=& - \tilde X_\HI
\int_{\nu_\HI}^{\infty} 4 \pi J_\nu \sigma_\HI {d\nu\over {\rm h}\nu}
\, ,
\label{Treioneq}
\end{eqnarray}
where we have used equation (\ref{P}) 
and have taken the crude approximation that
the only two processes which  significantly affect the temperature under the
short timescale of reionization (reciprocal of photoionization rate) are
photoionization heating of hydrogen:   
\begin{equation}
{dQ\over dt} \sim n_\HI \int_{\nu_\HI}^{\infty} 4 \pi
J_\nu \sigma_\HI ({\rm h} \nu - {\rm h} \nu_\HI) {d\nu\over {\rm
h}\nu}\, 
\label{qphoto}
\end{equation}
and the change in the total number of gas particles. Note that during this
brief period of radiation turn-on, the abundances of various species are
far from equilibrium.

To integrate the above equations, we adopt the further approximation that
the universe has only hydrogen, 
i.e.\ the term $\sum_i \tilde X_{i} = 2 - \tilde
X_\HI$. Putting $T = 0$ right before reionization because the temperature
right after is much higher, we obtain the large time limit of $T$:
\begin{equation}
T_{\rm reion} = {1\over {3} k_{\rm B}} E_J\, , \quad \quad E_J \equiv
\left[{\int_{\nu_\HI}^{\infty}
4 \pi J_\nu \sigma_\HI ({\rm h} \nu - {\rm h} \nu_\HI) {d\nu\over
{\rm h}\nu}} \, / \, \int_{\nu_\HI}^{\infty} 4 \pi J_\nu \sigma_\HI
{d\nu\over {\rm h}\nu} \right]\, ,
\label{Treion}
\end{equation}
where $E_J$ is kept constant after the radiation field is turned on.
This is the temperature that the intergalactic medium reaches asymptotically 
before recombination processes become important and halts the exponential
decrease 
of $\tilde X_\HI$, setting its value to that implied by ionization
equilibrium 
thereafter (hydrogen is now highly ionized). Making use of $J_{\nu} \propto
\nu^{-1}$, it can be shown that $T_{\rm reion} \sim 20000\dim{K}
$. This value is not significantly affected by the slope of $J_{\nu}$
assuming it is between $-1$ and $-1.5$ at frequencies immediately above
$\nu_\HI$, nor by the behavior of $J_{\nu}$ at still higher frequencies. 

Therefore, the first conclusion to draw is that for sudden reionization models,
the temperature of reionization is largely independent of $J_{\HI}$, density,
the epoch of reionization or even cosmology. We checked this for a few 
fluid elements in the computations shown in Fig.\ \ref{fig3abcd} 
and found that it
is indeed the case. One possible exception, which we will discuss in
\S~\ref{spectrum}, is that changing the spectrum of $J_\nu$ can affect the
helium abundance, which in turn has a non-negligible effect on $T_{\rm reion}$. 
For sufficiently high redshifts, $a_{\rm reion}\,\approxlt\,
1/15$, we also find $T_{\rm reion}$ to be a little less than the value given above,
because of non-negligible amount of Compton cooling.

Next, we consider the evolution of $T$ from the epoch of reionization to 
$z = 2, 3$ and $4$. Judging from the result in  Fig.\ \ref{fig3abcd} that 
for sufficiently early reionization, the temperature-density relation 
approaches an asymptote, it seems reasonable to assume that what matters is the
dominant thermal and chemical reaction rates at the later redshifts. 
In this regime, the dominant heating mechanism is again the photoionization of 
neutral hydrogen, aside from adiabatic heating/cooling (recall that we assume
a universe filled with only hydrogen).  
For sufficiently low temperature, recombination (of proton and electron)
cooling is subdominant compared to photoionization heating (at temperature of 
the order of $10^4\dim{K}$, recombination cooling rate is smaller 
but actually not negligible compared to photoionization heating; it
is again a crude approximation we adopt to make the problem tractable
analytically). Putting all these into
equation (\ref{T}), the thermal evolution is then approximated by:
\begin{equation}
{dT\over dt} = -2HT + {2T\over{3(1+\delta)}} {d\delta\over dt} + 
{2\over {3k_B}} {\tilde X_\HII \tilde X_{e} \over \sum_i \tilde X_{i}}
{\bar\rho_b\over m_p} (1+\delta) R E_J
\label{Tapprox1}
\end{equation}
where we have ignored the term in equation (\ref{T}) due to the change in 
the number of species, used equations (\ref{P}), (\ref{qphoto}) and
(\ref{Treion}) and assumed ionization  
equilibrium (setting the right hand side of eq. [\ref{n}] to zero), with 
$R$ approximately given by
\begin{equation}
R \approx 4 \times 10^{-13} ({T\over {10^4 {\rm K}}})^{-0.7} 
\dim{cm}^{3}\dim{s}^{-1}\, .
\label{R}
\end{equation}
A more accurate expression for $R$ is given in the Appendix and used
for our semi-analytical computation but this approximate form suffices
for the present analytical calculation.
Moreover, assuming again there is only hydrogen in the universe, we will
approximate 
${\tilde X_\HII \tilde X_{e} / \sum_i \tilde X_{i}}$ by $1/2$ (recall that
hydrogen is now highly ionized).

Even with this highly simplified equation, the problem of predicting
the temperature-density relation is still untractable analytically, if one
uses the density evolution as expressed in equation (\ref{zadelta}). 
One possible approximation is to assume the collapse/expansion occurs only
along one direction i.e. $1 + \delta = 1/ (1 - D_+ \lambda)$ where $\lambda$
is a constant specified by initial conditions. Equation (\ref{Tapprox1}) can
then be solved analytically but it turns out the resulting equations of 
state differ qualitatively from the ones shown in Fig.\ \ref{fig3abcd} in 
the sense that at high and low densities, the predicted temperature tends to 
be higher than what it actually is. This is because in three dimensions, most
fluid elements that reach higher densities do so by having one axis
collapsing while the two others expand. By ignoring the expansion in
the other two axes (as in a one-dimensional calculation) and thereby missing
extra adiabatic cooling, the temperature is systematically over-estimated. 
Moreover, most elements that reach the lower densities also have at least two
axes expanding and so the one-dimensional calculation again underestimates the
amount of cooling. The combination of both effects conspire to give
an equation of state that resembles a power-law for genuine three-dimensional
fluid elements.

The next best question to ask is then: given that the outcome is a power-law
equation of state, can we at least predict its amplitude and slope?
To do so, we turn to linear theory. We expand the logarithm of $T$ to
first order in $\delta$:
\begin{equation}
{\rm ln} T = {\rm ln} T_0 + (\gamma-1) \delta = {\rm ln} T_0 + (\gamma-1) [{\rm
ln}(1+\delta)] \, ,
\label{Texpand}
\end{equation}
where $T_0$ is simply the temperature of a fluid element that remains at
the cosmic mean density and $\gamma - 1$ multiplied by $\delta$ gives the
fluctuation about the mean (the notation $\gamma -1$ is chosen to agree with
the common notation as in eq. [\ref{eos}]). The second equality holds in the
small $\delta$ 
limit. The power-law slope relating temperature and density
for small $\delta$ is then given by $d \, {\rm ln}T /d \, {\rm ln} (1+\delta) =
\gamma - 1$.  
Knowing that a single power-law equation of state approximately holds
even for $\delta$ not very small, $\gamma - 1$ is then taken to be the correct
power-law slope throughout our range of interest i.e.
\begin{equation}
T = T_0 (1+\delta)^{\gamma-1}
\label{eos}
\end{equation}

Putting $\delta = 0$ into equation (\ref{Tapprox1}) and making use of the
relation $H = H_0 \Omeganow^{1/2} (1+z)^{3/2}$ where $\Omeganow$ is the ratio
of matter density to critical density today (the expression is
exact for $\Omeganow = 1$ and approximately true at sufficiently high redshift;
see Peebles 1993 Chapter 5), it is possible to show that
\begin{equation}
{d a^2 T_0 \over da} = T_0^{-0.7} a^{-{1\over 2}} B \,
\label{T0eq}
\end{equation}
where $B$ is defined by
\begin{equation}
B \equiv {\rho_{c100}\over m_p}  {T_{\rm reion}}
RT_0^{0.7} {\Omega_b 
h^2 \over {H_0 \sqrt\Omeganow}} \approx 10 \dim{K}^{0.7} T_{\rm
reion} 
\left[{1\over {h \sqrt\Omeganow}} {\Omega_b h^2 \over
0.0125}\right]\, 
\label{B}
\end{equation}
where $\rho_{c100}$ is the critical density today if $H_0 = 100\dim{km}
\dim{s}^{-1}\dim{Mpc}^{-1}$, 
$T_{\rm reion}$ is given by equation (\ref{Treion}) 
and $R$ is the recombination
rate give in equation (\ref{R}) (note that the combination $R T^{0.7}$ is
independent of temperature) and $\Omega_b$, $\Omeganow$, $h$ and $H_0$ are
cosmological parameters as explained in the Introduction. 
Note also temperature is measured in Kelvin and $B$ has the dimension of
${\rm K}^{1.7}$, hence the factor of ${\rm K}^{0.7}$ in the last expression.

Integrating equation (\ref{T0eq}), we obtain:
\begin{equation}
 T_0^{1.7} = ({a^2_{\rm reion}\over a^2} T_{\rm reion})^{1.7} + {1.7\over 1.9}
 a^{-{3\over 
 2}} \left[1-\left({a_{\rm reion}\over a}\right)^{1.9}\right] B \, .
\label{T0}
\end{equation}
The initial condition is chosen such that $T = T_{\rm reion}$
(eq. [\ref{Treion}]) at $a = a_{\rm
reion}$, the epoch of reionization.

The equation of evolution for $\gamma -1$ can be derived from
equation (\ref{Tapprox1}) by 
expanding $T$ as in equation (\ref{Texpand}) and collecting terms that are
first order in $\delta$:
\begin{equation}
{d (\gamma-1) \over da} + (\gamma-1) \left[ {\Omeganow^{0.6}\over a} + 1.7 {d \,
{\rm 
ln} a^2  
T_0\over da}\right] = {d {\rm ln} a^2 T_0\over da} + {2\over 3}
{\Omeganow^{0.6}\over 
a} \, . 
\label{T1eq}
\end{equation}
In deriving the above, we have made use of the fact that the linear growth 
factor $D_+$, which controls the growth of $\delta$ for small $\delta$ 
($\delta \propto D_+$) obeys the relation $d \, {\rm ln}D_+/ d\, {\rm
ln} a = \Omeganow^{0.6}$, which is exact for critical matter density and
approximate otherwise (Peebles 1980). 

Equation (\ref{T1eq}) is 
a linear ordinary differential equation for $\gamma$, which is straightforward,
if somewhat tedious, to solve. Recalling that $T_{\rm reion}$
is independent of $\delta$, we use the initial condition
$\gamma - 1 = 0$ at $a = a_{\rm
reion}$ and obtain:
\begin{equation}
\gamma - 1 = {1\over 1.7}\left[1 - \left(a_{\rm reion}^2 T_{\rm reion}\over {a^2
T_0}\right)^{1.7} \left({a_{\rm reion} \over
a}\right)^{{\Omeganow^{0.6}}}\right] + 
\left[{2\over 3} - {1\over 1.7}\right] (a^2 T_0)^{-1.7} \Omeganow^{0.6} C \, ,
\label{T1}
\end{equation}
where $C$ is defined by
\begin{eqnarray}
C \equiv && {1.7\over 1.9} B \left[{a^{1.9}\over{1.9+\Omeganow^{0.6}}}
\left(1-({a_{\rm reion}\over a})^{1.9+{\Omeganow^{0.6}}}\right)-
{a_{\rm reion}^{1.9}\over \Omeganow^{0.6}} \left(1-({a_{\rm reion}\over
a})^{\Omeganow^{0.6}}\right)\right] \\ \nonumber
&& + {(a_{\rm reion}^2 T_{\rm reion})^{1.7}\over \Omeganow^{0.6}} \left[1 -
({a_{\rm 
reion}\over a})^{\Omeganow^{0.6}}\right] \, 
\label{C}
\end{eqnarray}
where $B$ and $T_0$ are defined in equations (\ref{B}) and (\ref{T0}). 
The slope $\gamma-1 = d \, {\rm ln}T /d \, {\rm ln} (1+\delta)$ exhibits the
correct qualitative behavior found using our semi-analytical computation shown
in Fig.\ \ref{fig3abcd}. 
We plot in Fig.\ \ref{gamma} the evolution of $\gamma -1$ as given by
equation (\ref{T1}) for a number of different $a_{\rm reion}$. 
An interesting consequence of the above expression for the slope is
a limit as to how steep the temperature-density relation can be 
(or an upper limit to $\gamma-1$). In the limit where reionization occurs very
early: 
\begin{equation}
\gamma_{\rm max} - 1 = {1\over 1.7} + {{0.078 \Omeganow \over {1.9 +
\Omeganow^{0.6}}}} \, .
\label{T1limit}
\end{equation}
Hence, the maximum possible value for $\gamma -1$ is 0.62 (assuming 
$\Omeganow \, \leq \, 1$). This has
interesting 
implications for the slope of the column density distribution of the Lyman
alpha forest, which we will discuss in \S~\ref{discuss}.

\def\capgamma{
The evolution of $\gamma -1$ as given by equation (\protect{\ref{T1}}) for 
$a_{\rm reion} = 1/6$ ({\it solid line}), $a_{\rm reion} = 1/8$ ({\it dotted
line}) and $a_{\rm reion} = 1/11$ ({\it dashed line}). Here
$T_{\rm reion} = 20000 \, 
{\rm K}$  and $\Omeganow = 1$ in all cases.
}
\begin{figure}
\par\centerline{%
\epsfxsize=0.5\columnwidth\epsfbox{\figdir/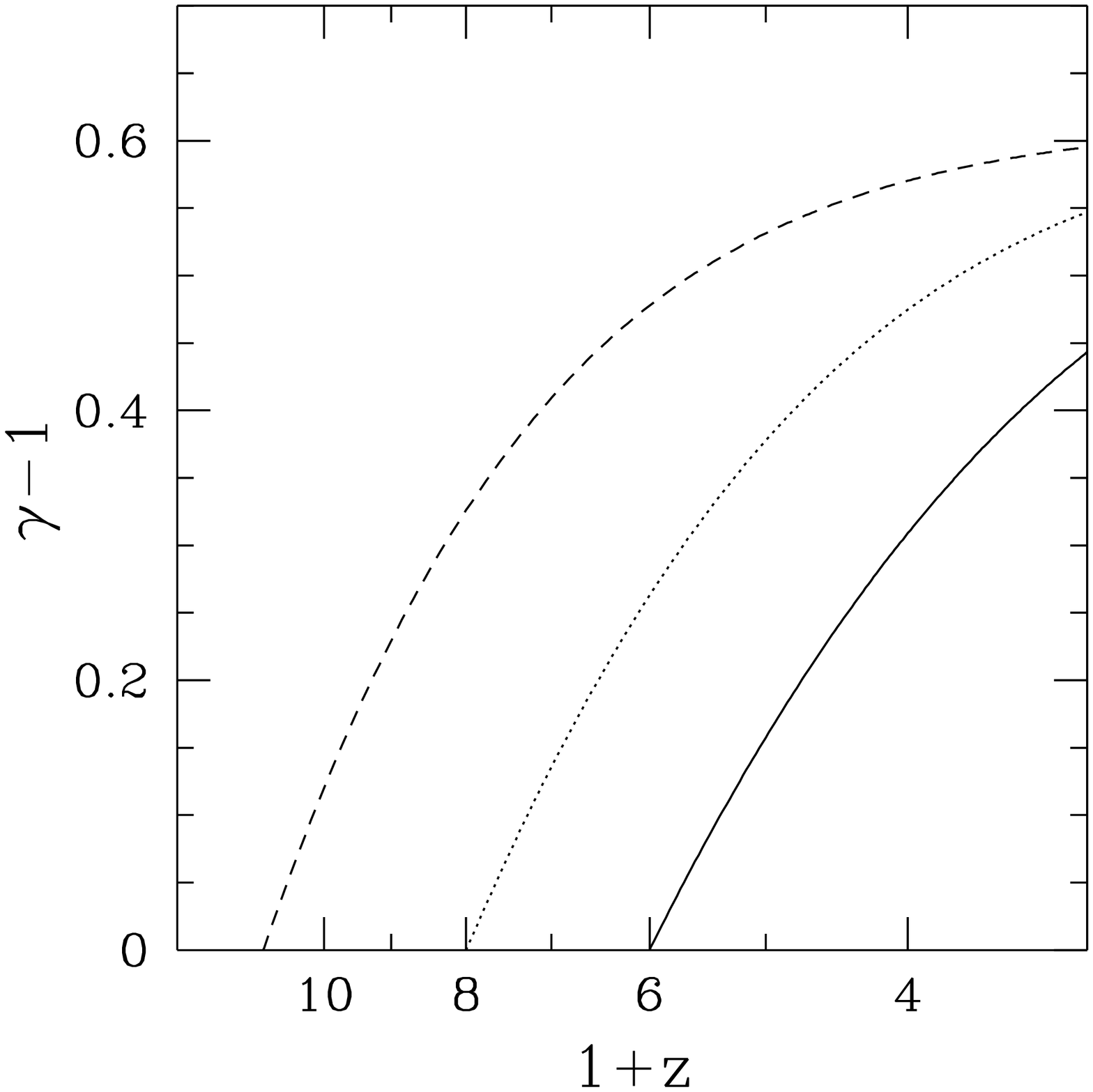}}%
\caption{\capgamma}
\label{gamma}
\end{figure}

\def\capEab{
Temperature-density relation for two sudden-reionization models:
comparison of results of semi-analytical calculations (black dots) and
predictions based on analytical approximations expressed in
equations (\protect{\ref{Texpand}}), (\protect{\ref{T0}}),
 and (\protect{\ref{T1}})
({\it solid, dotted and dashed lines}). The model parameters are described in
the  
caption of Fig.\ \protect{\ref{fig3abcd}}. 
The black dots in the left panel and right panel
are exactly the same points as those in Fig.\ \protect{\ref{fig3abcd}}a 
and Fig.\ \protect{\ref{fig3abcd}}c respectively , i.e.\ sudden
reionization at  $z = 6$ and $z = 19$. From top to bottom for each figure, the
results are shown at $z = 4$, $3$, and $2$ respectively for the black dots
and also for the lines. Notice how the predicted slopes match quite well
while the amplitudes are off. A modified version of equation (\protect{\ref{T0}})
(eq. [\protect{\ref{T0modified}}]) gives much better fit to the amplitudes
and is used instead in the rest of this paper.
}
\begin{figure}
\par\centerline{%
\epsfxsize=0.5\columnwidth\epsfbox{\figdir/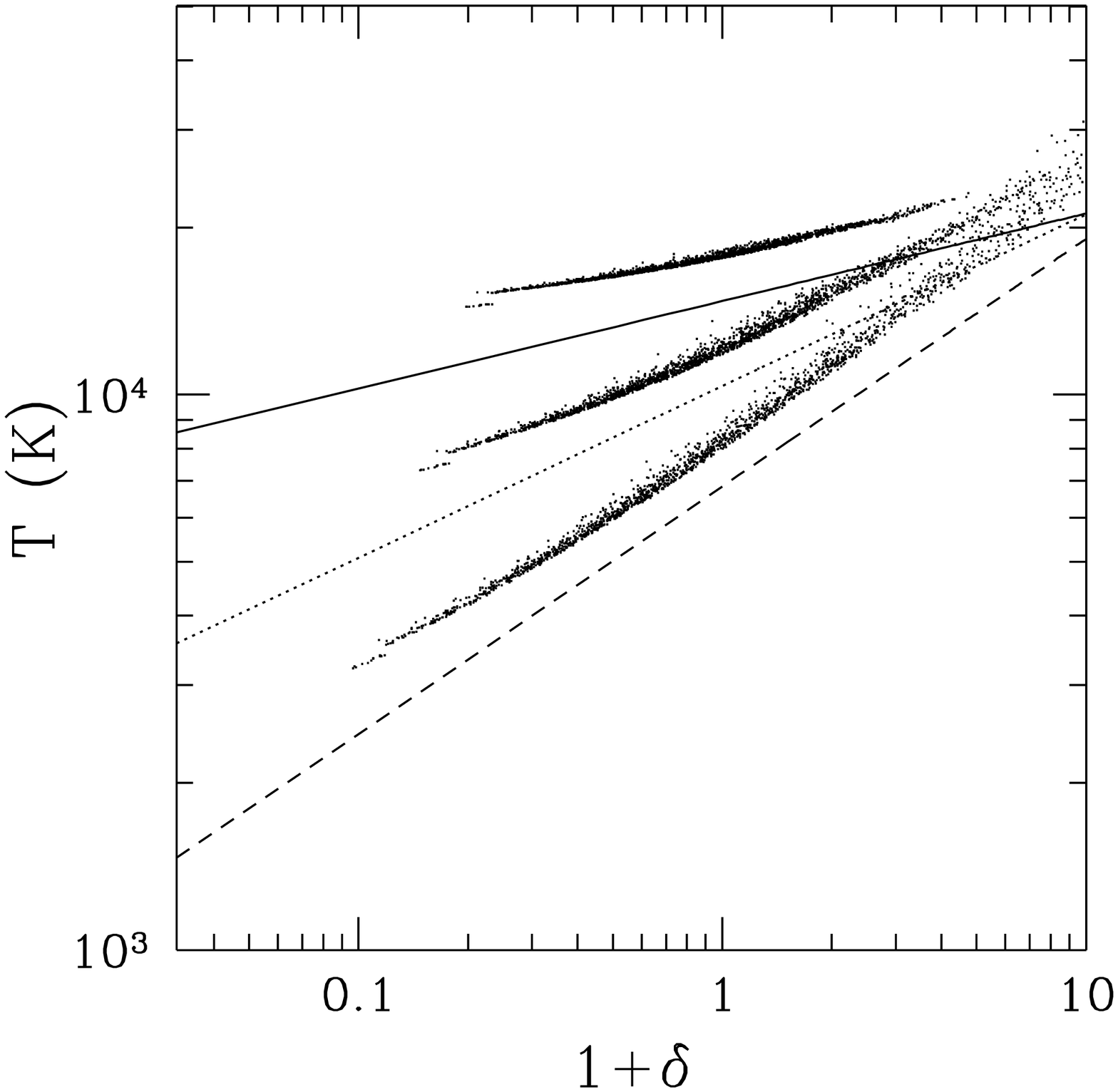}\hfil%
\epsfxsize=0.5\columnwidth\epsfbox{\figdir/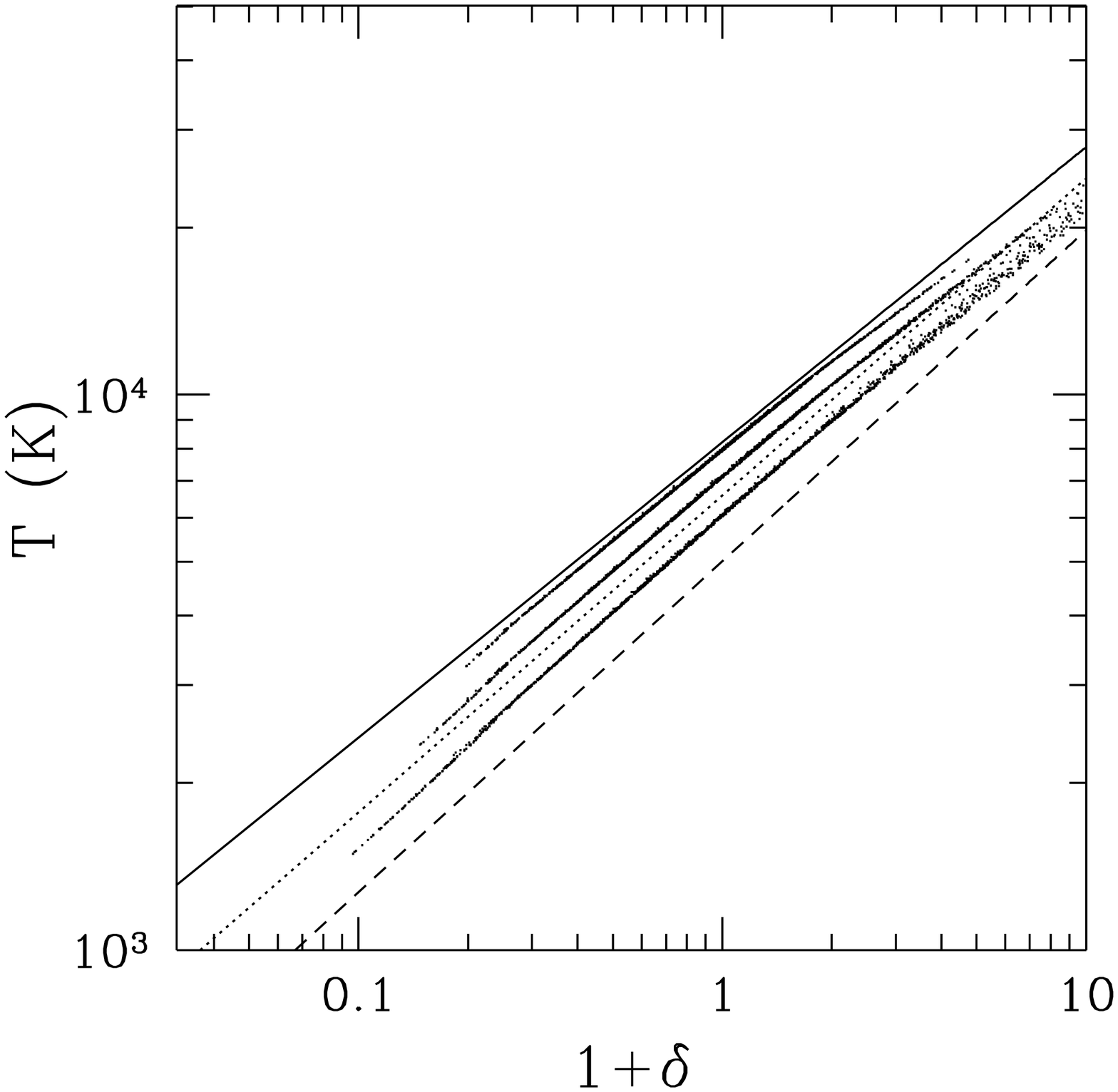}}%
\caption{\capEab}
\label{fig5ab}
\end{figure}

We show in Fig.\ \ref{fig5ab} a  comparison of the above expressions
(eq. [\ref{Texpand}], [\ref{T0}] and [\ref{T1}]) with the temperature-density
relation obtained by solving numerically the rate equations as outlined 
in Sec~\ref{equation}. The evolution of the slope is quite accurately captured
by our  analytical approximation in equation (\ref{T1}).
However, the amplitude $T_0$ predicted by equation (\ref{T0}) is off:
the overall predicted temperature is lower than what it actually is and the
the predicted spread in $T_0$ from $z = 2$ to $z = 4$ is also larger
than what the semi-analytical calculation shows (Fig.\ \ref{fig5ab}b).
This is mainly because helium is not taken into account at all in the above
treatment. By ignoring heating due to photoionization of helium, we
systematically underestimates $T_0$ in our analytical calculation.
The time evolution for $T_0$ is also not quite right for the same reason.
We therefore introduce the following modification of equation 
(\ref{T0}) that gives an
excellent fit to the results of the semi-analytical calculations shown in
Fig.\ \ref{fig3abcd}: 
\begin{equation}
T_0^{1.7} = ({a_{\rm reion}^2\over a^2} \tilde T_{\rm reion})^{1.7} + {1.7\over
 1.9} a^{-{3\over 
 2}+{{a-0.25}\over a}} \left[1-{a_{\rm reion}\over a}^{1.9}\right] B \, \quad ,
 \quad \tilde T_{\rm reion} = 25000 \dim{K}
\label{T0modified}
\end{equation}
where we have modified the power index of $a$ in the second term on the
right hand side and used $\tilde T_{\rm reion} = 25000$, in place of $T_{\rm
reion}$ given in equation (\ref{Treion}). In general, a different best-fit $\tilde
T_{\rm reion}$ would have to be chosen for different abundances of the helium 
species, when one varies the spectrum of the ionizing radiation, for instance.
We will return to this point later.

On the other hand, the slope of the equation of state $\gamma - 1$ is
quite insensitive to the exact value of $T_{\rm reion}$, and using
the original estimate (eq. [\ref{Treion}]) in equation (\ref{T1}) is sufficiently
accurate. 

The analytical predictions for the equation of state shown
in Fig.\ \ref{fig3abcd} 
are based on this modified version of $T_0$ (eq. [\ref{T0modified}])
together with equations (\ref{Texpand}) and (\ref{T1}). Note that, for
the case in which the universe reionizes at $z = 19$, the temperature-density
relation is very close to that of the case in which the universe reionizes
at $z = 10$. This turns out to be a general trend, that there is an asymptotic
temperature-density relation when the universe reionizes 
sufficiently early. To see how this asymptotic state is approached,
we show in Fig.\ \ref{asymptote} the evolution in temperature of one
particular fluid element for different reionization histories. Notice
how the solid line (sudden reionization at $z = 10$) and the dotted line
(sudden reionization at $z = 19$) converges. The other lines
in the figure corresponds to reionization models that will be discussed
later. We also find that fluid elements at other densities also exhibit the
same kind of convergence towards an asymptotic state.

The expressions in equations (\ref{Texpand}), (\ref{T1}), and
(\ref{T0modified}) are derived by ignoring processes like Compton cooling
which are important at higher redshifts, if the universe is already reionized
by  then. Hence, the analytical expressions are not quite as accurate 
for $a_{\rm reion} < 1/11$. However, we can make use of the knowledge of
asymptotic behavior for early reionization models and use simply 
the $a_{\rm reion} = 1/11$ result of equations (\ref{Texpand}), (\ref{T1}),
 and
(\ref{T0modified}) for any models in which reionization occurs earlier
than $z = 10$. This is what we have done in making Fig.\ \ref{fig3abcd}d.

We now turn to two other interesting variations of the sudden-reionization
models. 

\subsection{Variation with Cosmology}
\label{cosmology}

\def\capomegab{
The temperature-density relations for $\Omega_b h^2 = 0.025$ (upper black
dots) and $\Omega_b h^2 = 0.0125$ (lower black dots) at $z = 3$ for sudden
reionization model with radiation-turn-on at $z = 7$. The rest of the
cosmological parameters are exactly the same as those given in
Fig.\ \protect{\ref{fig3abcd}}. The dotted and solid lines are the predictions
according to equations (\ref{Texpand}), (\ref{T1}) and (\ref{T0modified}),
respectively for the two models.
}
\begin{figure}
\par\centerline{%
\epsfxsize=0.5\columnwidth\epsfbox{\figdir/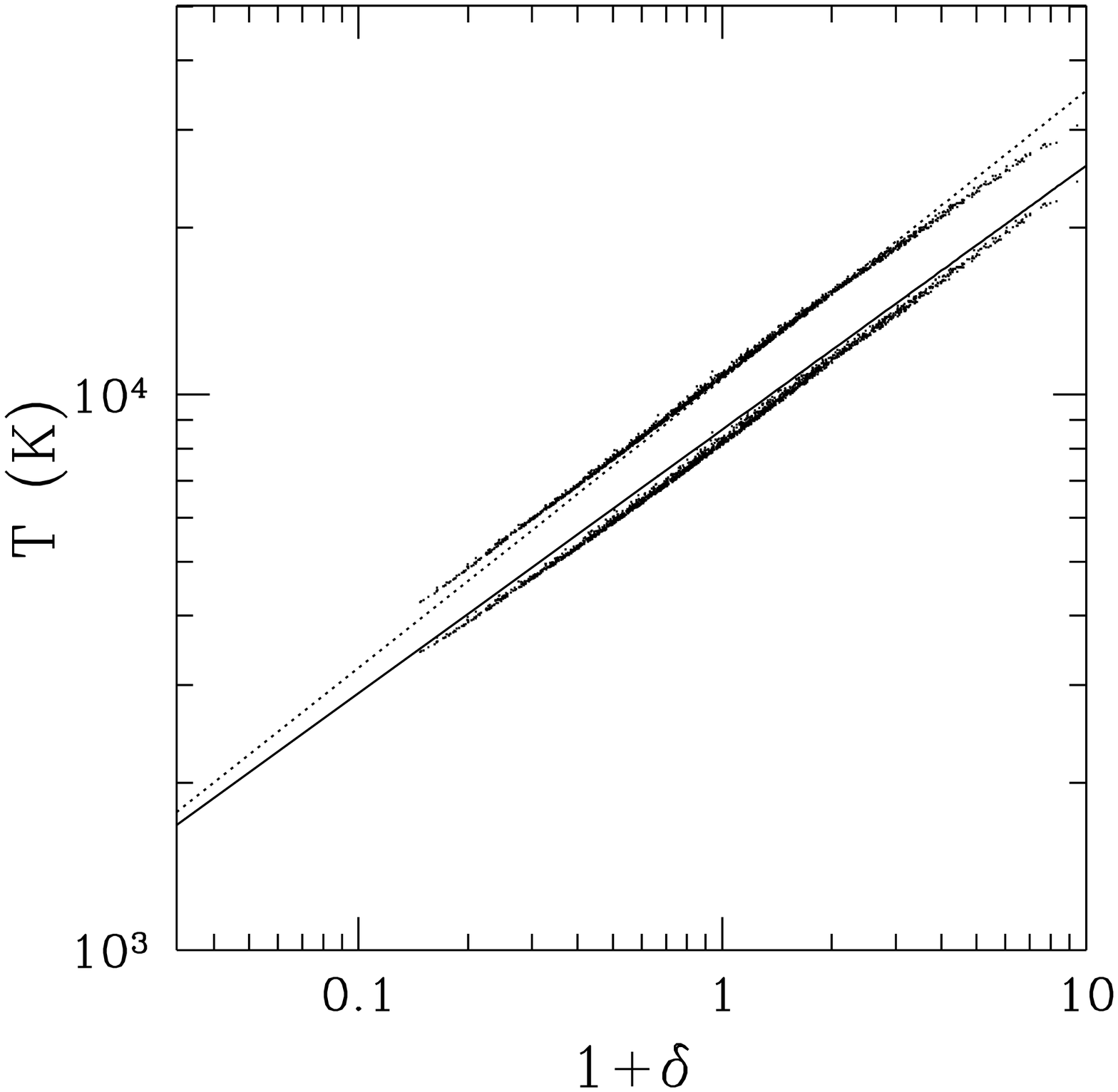}}%
\caption{\capomegab}
\label{omegab}
\end{figure}

We investigate the temperature-density relations for different cosmological
parameters, including 
the baryon density, the Hubble constant and the fraction of critical density
in matter. One example is shown in Fig.~\ref{omegab}, where $\Omega_b h^2$ has
been increased by a factor of two from the value in Fig.~\ref{fig3abcd}. 
Once again, we find that a power-law equation of state gives
an adequate description of the temperature-density relation. Moreover, the
analytical expressions in equations (\ref{Texpand}), (\ref{T1}), and
(\ref{T0modified}) work quite well in describing the change in amplitude and
slope. As predicted by equations (\ref{T0modified}) and (\ref{B}), the overall
temperature increases as $\Omega_b h^2$ is raised because the rate of
photoionization heating is increased. (Recombination cooling rate is actually
also increased but photoionization heating is the dominant entropy changing
process.)  

We have also tried other variations such as lowering $\Omeganow$.
A realistic structure formation model such as the cosmological constant
dominated Cold Dark Matter model with $\Omeganow = 0.35$ and $h = 0.7$ (Kofman,
Gnedin, \& Bahcall 1993; Ostriker \& Steinhardt 1995) 
does not give a very different temperature-density
relation from the ones shown in Fig.\ \ref{fig3abcd} (using the same $\Omega_b
h^2$). This can be understood from equations (\ref{T0modified}) and (\ref{B}),
where 
it is clear the overall 
temperature to the $1/1.7$-th power is dependent upon the combination $h
\sqrt\Omeganow$ which only varies from $0.5$ for the model in Fig.\
\ref{fig3abcd} to about $0.4$ in the cosmological constant model.  

In general, we find that equations (\ref{Texpand}), (\ref{T1}) and
(\ref{T0modified}) give an agreement to the actual temperature-density relation
to within $20 \%$ for a variety of cosmological models we have tested.
A large part of this error has to do with the fact that changing $\Omeganow$
and $h$ changes the amount of time helium photoionization heating or
recombination cooling has to act to make a difference to the thermal evolution.
The effect of helium is what we will discuss next.

\subsection{Variation with Amplitude and Spectrum of Radiation}
\label{spectrum}

\def\capFab{
The effect of varying the amplitude of $J_\nu$ on the
temperature-density relation. Black points in the left panel (Fig.\
\protect{\ref{amplitude_fig7}}a) are the
result of 
semi-analytical calculation of sudden reionization at $a_{\rm reion} = 1/8$
with $J_{\rm ion} = 0.1$ (eq. [\protect{\ref{Jevol}}]) while those 
in the right panel (Fig.\ \protect{\ref{amplitude_fig7}}b) 
has $J_{\rm ion} = 2.0$ starting from the same
$a_{\rm reion}$. The spectrum is chosen to be the same with $f = 0.01$
(eq. [\protect{\ref{cutoff}}]) for both
panels. The outputs from top to bottom are at $z = 4, 3$ and $2$.
The same solid, dotted, and dashed lines are drawn for
both panels to facilitate comparison. They are also exactly the same as
the lines in Fig.\ \protect{\ref{fig3abcd}}b, 
based on the analytical expressions in 
equations (\protect{\ref{Texpand}}), (\protect{\ref{T1}}), and
(\protect{\ref{T0modified}}), using $a_{\rm reion} = 1/8$.
The cosmological parameters in Fig.\ \protect{\ref{amplitude_fig7}} are exactly the
same as those in Fig.\ \protect{\ref{fig3abcd}}.
}
\begin{figure}
\par\centerline{%
\epsfxsize=0.5\columnwidth\epsfbox{\figdir/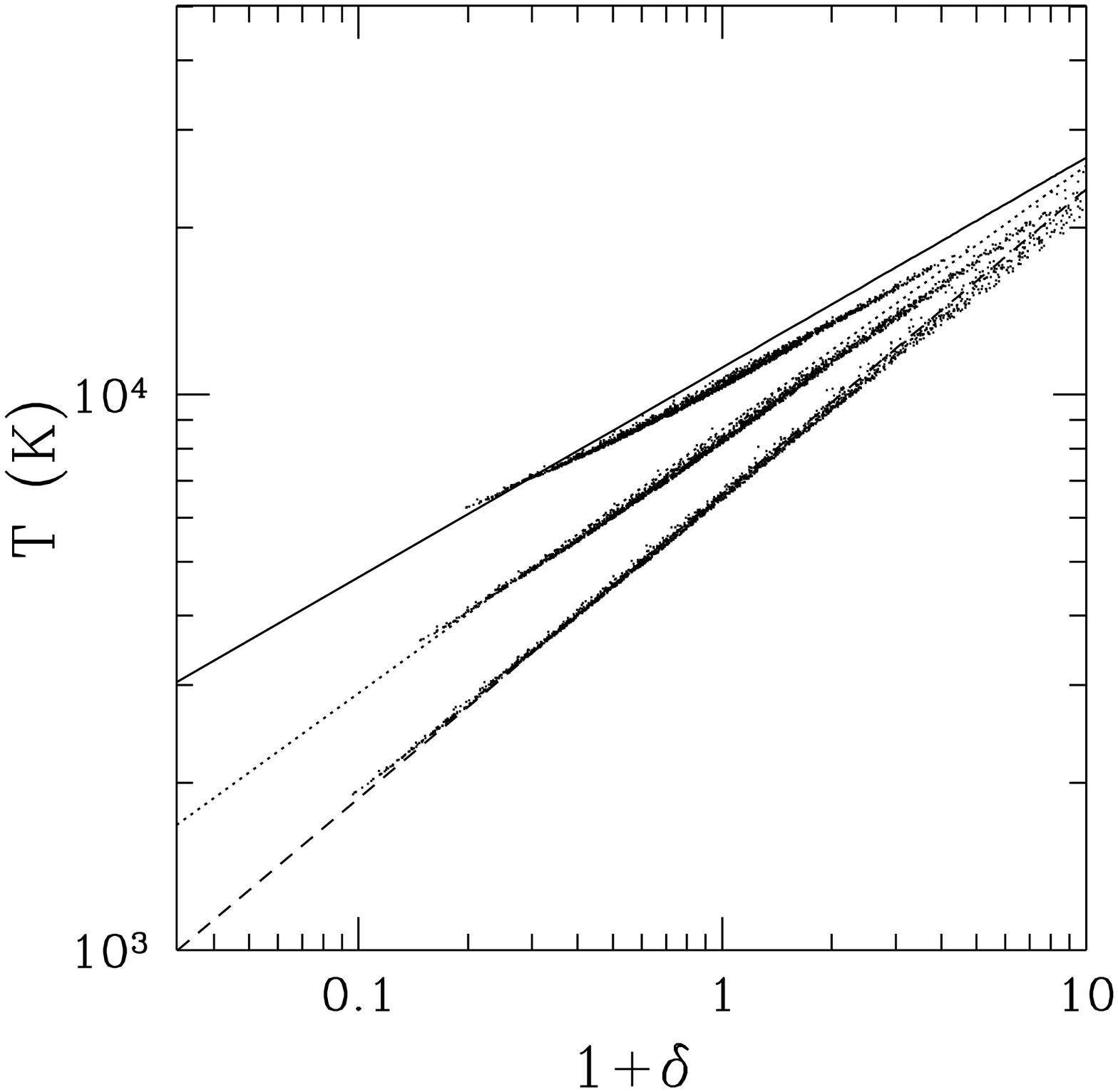}\hfil%
\epsfxsize=0.5\columnwidth\epsfbox{\figdir/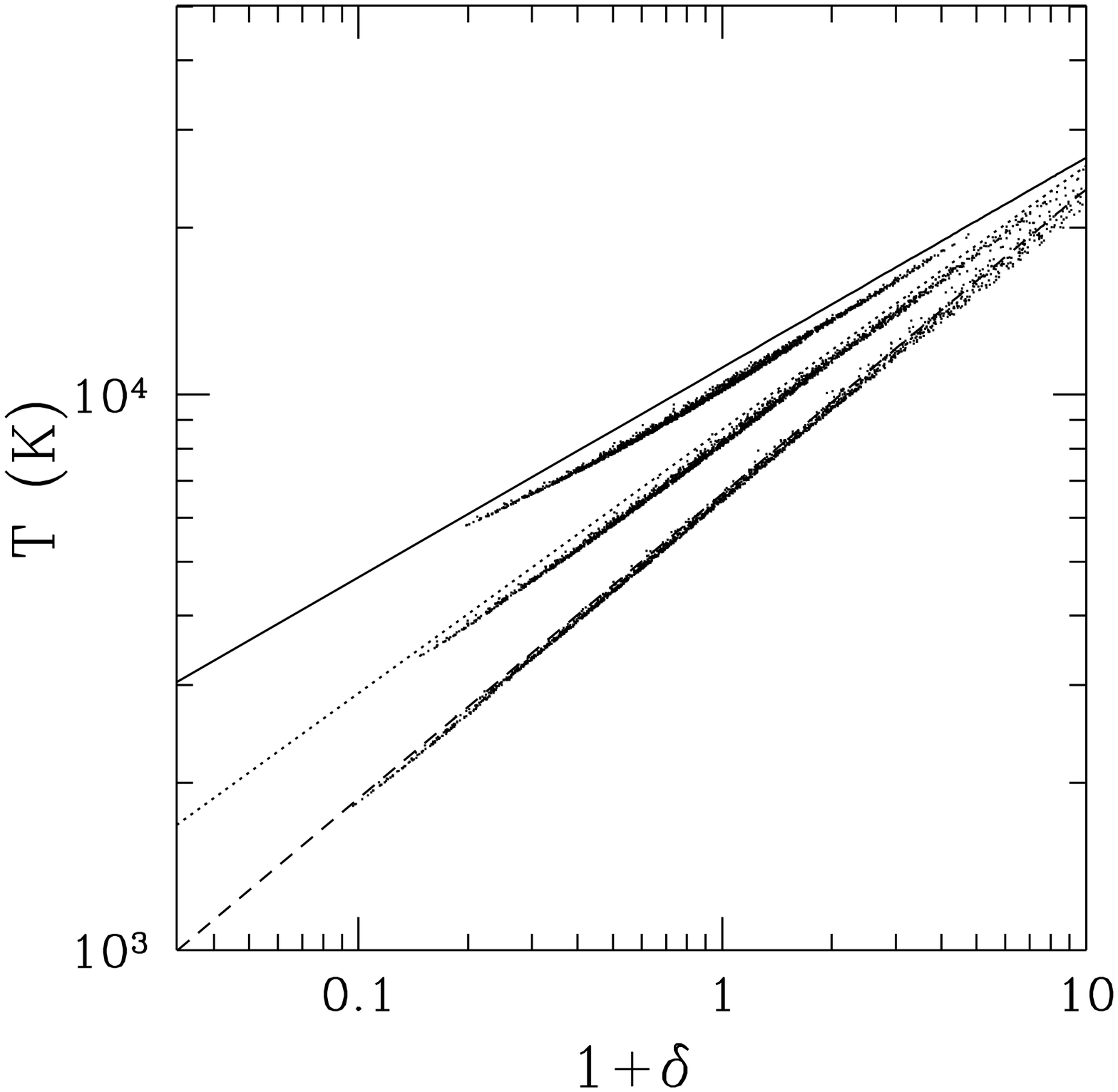}}%
\caption{\capFab}
\label{amplitude_fig7}
\end{figure}

The last set of tests we perform for the class of sudden reionization models is
to vary the spectrum and amplitude of the ionization radiation. 
We find that varying the amplitude of the ionization radiation does not
affect the temperature-density relation significantly, as predicted by
equations (\ref{T1}) and (\ref{T0modified}), as long as the flux level is
sufficient to maintain the universe highly ionized. This is shown in Fig.\ 
\ref{amplitude_fig7}a
and \ref{amplitude_fig7}b, which corresponds 
to choosing $J_{\rm ion} = 0.1$ and $J_{\rm ion} = 2.0$
respectively in equation (\ref{Jevol}).
A comparison of the two figures reveals a slight decrease in
temperature when $J_\HI$ is raised, 
which is a result of more opportunities for recombination cooling, but
the effect is very small and becomes insignificant at lower
redshifts ($z = 2$).

\def\capGab{
The effect of varying radiation spectrum on the temperature-density
relation. Black dots are the results obtained by the semi-analytical
calculation explained in \S~\ref{equation}. The reionization and cosmological
model here is, except for the spectrum, exactly the same as 
that in Fig.\ \protect{\ref{fig3abcd}}a 
where the universe reionizes suddenly at $z = 6$.
The {\it solid, dotted} and {\it dashed lines} of Fig.\ 
\protect{\ref{fig3abcd}}a are reproduced here
for comparison. Above and below the solid line at the top are two sets of points
at $z = 4$ which correspond to choosing $f = 1.0$ (less $\GII$) and $f =
10^{-4}$ (more $\GII$) respectively in
equation (\protect{\ref{cutoff}}). The same is also true for the two sets of points
surrounding the dotted line, 
except that they are at $z = 3$. Results are not shown for $z = 2$ because
the two different $f$'s give significantly overlapping results that are 
well-represented by the dashed line. The amplitude $J_\HI = 0.5$
(eq. [\protect{\ref{JHI}}]) is adopted throughout.
}
\begin{figure}
\par\centerline{%
\epsfxsize=0.5\columnwidth\epsfbox{\figdir/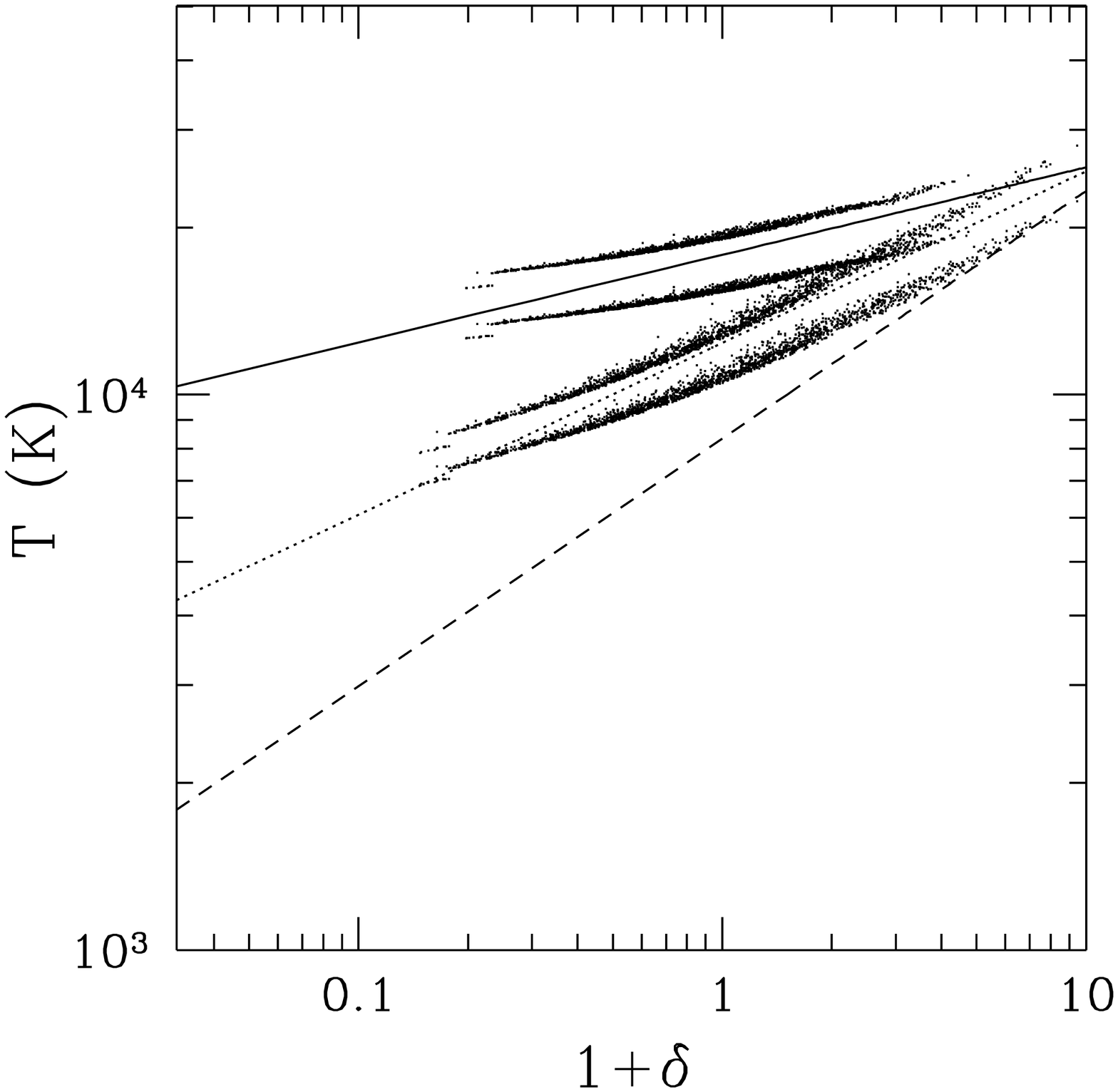}}%
\caption{\capGab}
\label{spectrum_fig8}
\end{figure}
The main effect of varying
the spectrum (when fixing $J_\HI$ defined in eq. [\ref{JHI}]) is  
that one changes the relative abundance of the different species of helium. 
which could have a non-negligible effect on the temperature due to 
recombination
cooling of the ionized species and photoionization heating of neutral helium. 
We test this effect by varying the $\GII$ cut-off factor $f$ in 
equation (\ref{cutoff}). 
Two examples are shown in Fig.\ \ref{spectrum_fig8} where we replot the lines in
Fig.\ \ref{fig3abcd}a 
(which match the temperature-density relations for $f = 0.01$ quite
well) together with the result of two semi-analytical calculations
for $f = 1.00$ and $f = 10^{-4}$ respectively.
By decreasing $f$ or the amount of radiation energetic enough to ionize $\GII$,
one increases the abundance of $\GII$ offering more opportunities for
recombination cooling, thereby lowering the overall temperature.
The reverse happens when $f$ is raised. The change in temperature due to this
effect could be as large as $20 \%$, with less change as one goes to lower
redshift, for a fixed reionization history. One can compensate
the error in the analytical estimate of $T_0$ (which ignores the effect
of helium) by lowering/raising $\tilde T_{\rm
reion}$ in equation (\ref{T0modified}) as one decreases/increases $f$ (For
instance, choosing
$\tilde T_{\rm reion} = 28000$ for the $f = 1$ case and $\tilde T_{\rm reion} =
22000$ for the $f = 10^{-4}$ works reasonably well).

The power-law slope, on the other hand, does not 
vary significantly with the spectrum of the ionizing background. It can be seen
from Fig.\ \ref{spectrum_fig8} 
that there is a slight decrease of slope for the corresponding
redshifts as one lowers $f$. This is due to the fact that $\GII$
recombination cooling has a stronger effect at higher density. The analytical
prediction for the slope in equation (\ref{T1}) works best when the effects of
helium is negligible, as can be seen from the better match of slopes for
the $f = 1$ case. Nevertheless, on the whole, we find that the prediction of
equation (\ref{T1}) for the slope is quite robust. 

To summarize the results of this section, we find that the temperature-density
relation at $z$ of $2$ to $4$ for sudden-reionization models is well described
by a power-law 
equation of state, with a scatter that depends on the total number of
fluid elements. The overall amplitude decreases and the slope steepens
as the epoch of reionization is pushed earlier; but the equation of state
rapidly approaches an asymptotic limit for reionization occurring at 
redshifts beyond $10$. 
Quantitative estimates for the slope  and amplitude
are given in equations (\ref{T1}) and (\ref{T0modified}). While
the analytical prediction for the slope is relatively robust for
various cosmologies and spectra, the prediction for
the amplitude agrees with the results of the semi-analytical calculations
(outlined in \S~\ref{equation})
to within $20 \%$ for reasonable range of cosmological and ionization
parameters. Most of the error comes from the neglect of helium in obtaining the
analytical estimate, which can be compensated if one allows 
$\tilde T_{\rm reion}$ to change in equation (\ref{T0modified}) according to
the abundance of 
the helium species (Fig. \ref{spectrum_fig8}).

\section{Reionization Preceded by Reheating}
\label{preheat}

The idealization of sudden reionization in the last section allows us to
understand the various trends of the temperature-density relation analytically. 
It is quite possible that the ionizing radiation is turned on more gradually in
nature.
In particular, we focus on a class of models in which 
radiation is turned on at some early time, with an intensity 
large enough to heat up the intergalactic medium but
small enough so
that the neutral fractions remain high (large recombination rates) though
slowly decreasing. 
This is motivated by hydrodynamic simulations (see Gnedin \& Ostriker 1996) in
which collapsed regions of high density (could be stars/quasars) are assumed to
give out radiation. 
As the formation rate of these regions increases with time and 
the neutral fraction drops steadily, the ionizing intensity climbs until
at some point, the universe becomes highly ionized (neutral fraction of
hydrogen less than $10^{-4}$), thus completing the reionization process when
the radiation levels off and species abundances are simply given by
ionization equilibrium. The simulation shown in Fig.\ \ref{fig1ab}a and 
\ref{fig2ab} is one such
example. 

To investigate the implications of these reionization models systematically, 
we adopt the following parametrization for the evolution of the ionizing
radiation:
\begin{equation}
\label{Jpreheat}
J_\HI = \cases{
J_{\rm ion} & for $a \geq a_{\rm ion}$, \cr
J_{\rm heat} \left({a\over a_{\rm ion}}\right)^7 &
for $a_{\rm ion} > a \geq a_{\rm heat}$,\cr
0 & for $a_{\rm heat} > a $,}
\end{equation}
where $J_\HI$ is defined in equation (\ref{JHI}), $a_{\rm heat}$ and $a_{\rm
ion}$ denote the size of the Hubble scale factor at the onset of reheating
and reionization. The slope $a^7$ in the second row of equation
(\ref{Jpreheat}) comes from fitting $J_\HI(a)$ in Fig.\ \ref{fig2ab}.
The spectrum of $J_\nu$ is assumed to have the same
form as in equation (\ref{cutoff}) with $f = 0.01$ and $J_0$ chosen to match
whatever $J_\HI$ specified above.

Before discussing the results for these models, let us point out that
in Fig.\ \ref{fig1ab}, 
we have already shown an example of the temperature-density relation
of a reionization-preceded-by-reheating model where $J$ varies with time as
in Fig.\ \ref{fig2ab}. 
Notice how the temperature-density relation can be approximated
by an equation of state resulting from a sudden reionization model (the solid
line in Fig.\ \ref{fig1ab}). 
The same turns out to be true for all other reionization
models we have tested, with the parametrization given in 
equation (\ref{Jpreheat}).

\def\capHab{
Black dots in Fig.\ \protect{\ref{preheat_fig9}}a on the left 
show the temperature-density
relation for a reionization model in which $a_{\rm heat} =
1/20$, $a_{\rm ion} = 1/6$, $J_{\rm heat} = 0.001$ and $J_{\rm ion} = 0.5$ (see
eq. [\protect{\ref{Jpreheat}}]). 
The three sets of points are from
top to bottom at $z = 4, 3$ and $2$. The solid, dotted and dashed 
lines are analytical approximations of sudden reionization models expressed 
in equations (\protect{\ref{Texpand}}), (\protect{\ref{T1}}), and
(\protect{\ref{T0modified}}), choosing $a_{\rm reion} = 1/8.4$.
Fig.\ \protect{\ref{preheat_fig9}}b on the right, 
on the other hand, corresponds to the choice of
parameters $a_{\rm heat} = 1/20$, $a_{\rm ion} = 1/6$, $J_{\rm heat} = 0.1$ and
$J_{\rm ion} = 0.5$ in  
eq. [\protect{\ref{Jpreheat}}]. The analytical approximations shown have
$a_{\rm reion}$ equal to $1/10$ in equation (\protect{\ref{Texpand}}),
(\protect{\ref{T1}}) and (\protect{\ref{T0modified}}). 
For both Fig.\ \protect{\ref{preheat_fig9}}a and \protect{\ref{preheat_fig9}}b,
the cosmological model is the same as that in Fig.\ \protect{\ref{fig3abcd}}.}
\begin{figure}
\par\centerline{%
\epsfxsize=0.5\columnwidth\epsfbox{\figdir/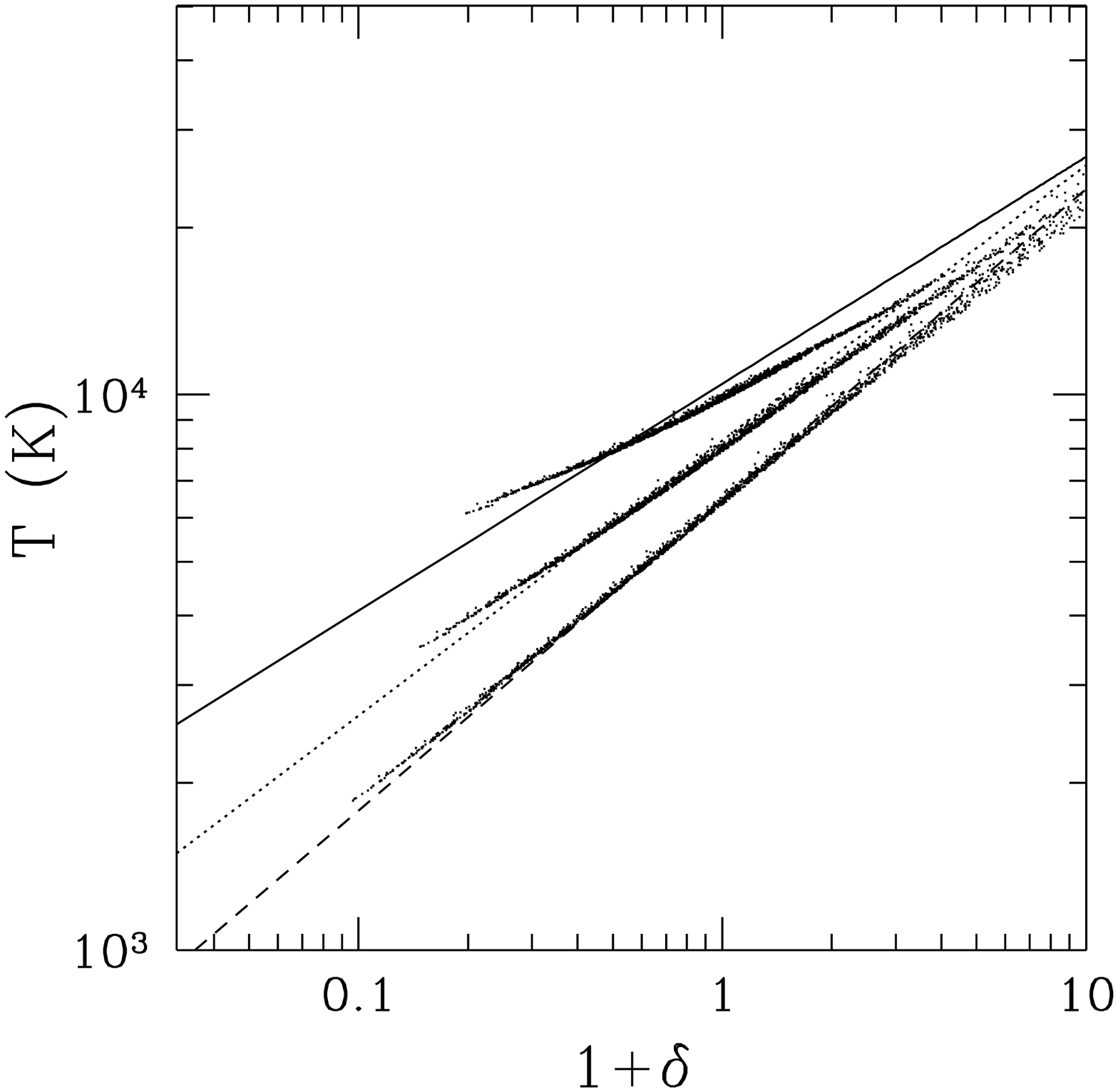}\hfil%
\epsfxsize=0.5\columnwidth\epsfbox{\figdir/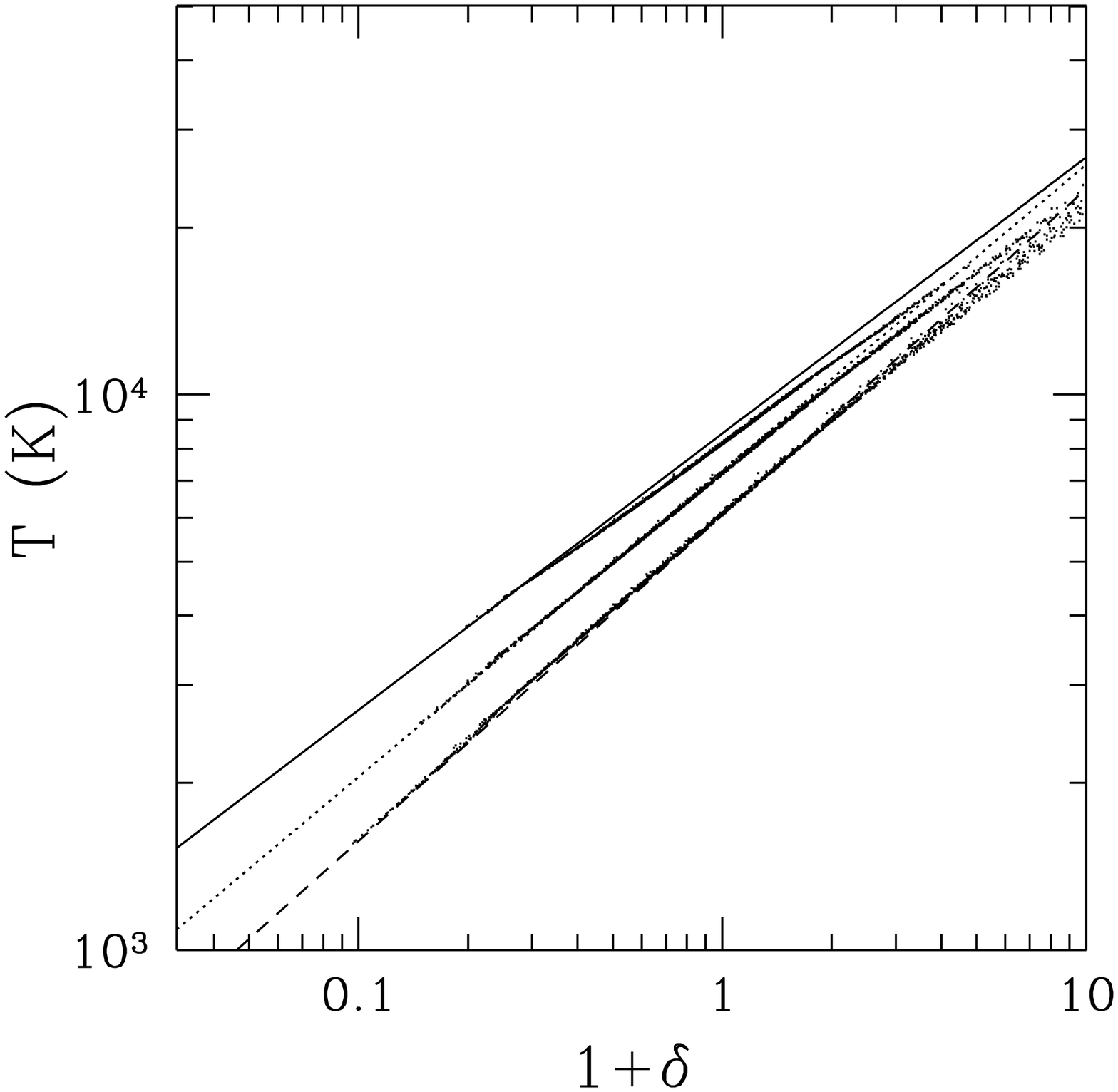}}%
\caption{\capHab}
\label{preheat_fig9}
\end{figure}

Two interesting examples are shown in Fig.\ \ref{preheat_fig9}. 
Both have the same epochs of reheating and reionization ($a_{\rm heat}$ and
$a_{\rm ion}$). The only difference is the amount of reheating that occurs:
$J_{\rm heat}$ being $0.001$ versus $0.1$ in equation (\ref{Jpreheat}). 
They can both be approximately fitted with sudden reionization models (no
preceding reheating period) 
where the epoch of sudden reionization is chosen to lie somewhere between
$a_{\rm heat}$ and $a_{\rm ion}$. The larger the amount of reheating, the
earlier the epoch of reionization is required. It is also interesting to
note that in Fig.\ \ref{preheat_fig9}a, 
the fit to the result at $z = 4$ is not as good 
as those at $z = 3$ or $z = 2$. It is because 
it takes some time for the universe to settle to its asymptotic
state (i.e a state which is independent of prior history, whether it be
sudden reionization, gradual reionization with preheating, etc).
The same should also be true for Fig.\ \ref{preheat_fig9}b, 
except that because the amount of
preheating is more significant, the universe, in effect, has more time
to settle to its asymptotic state, hence the better agreement with
sudden reionization models for all three redshifts.

To illustrate this convergence towards an asymptotic state,  
we show in Fig.\ \ref{asymptote} 
the evolution of temperature for one particular fluid
element for four different reionization histories. The tendency to
approach the same limiting temperature, for a given density evolution, can be
clearly seen. Note how the model illustrated with a long-dashed line evolves
to a temperature a little higher than the others, by $z = 2$. This is
because of its relatively small amount of reheating (small $J_{\rm heat}$ in
eq. [\protect{\ref{Jpreheat}}]) which implies the universe effectively
reionizes at a latter redshift, giving it less time to settle to its
asymptotic state.

\def\capIab{
Examples of convergence towards an asymptotic state. In the upper panel
is shown the evolution of the overdensity of a fluid element as a function of
redshift. In the lower panel is shown the evolution of temperature for the
same fluid element, but for a variety of reionization histories. 
The solid and dotted lines both correspond to sudden reionization
models in which $a_{\rm reion} = 1/11$ and $a_{\rm reion} = 1/20$
respectively and $J_{\rm ion} = 0.5$ in equation (\protect{\ref{Jevol}}). Their
reionization and cosmological parameters are exactly the same as those in
Fig.\ \protect{\ref{fig3abcd}}c and \protect{\ref{fig3abcd}}d respectively. 
The short dashed and long dashed lines
correspond exactly to the reionization models shown in 
Fig.\ \protect{\ref{preheat_fig9}}b and \protect{\ref{preheat_fig9}}a
respectively i.e. reionization-preceded-by-reheating models with $a_{\rm heat} =
1/20$, $a_{\rm ion} = 1/6$, $J_{\rm ion} = 0.5$ and
$J_{\rm heat} = 0.1$ for short dashed line and $J_{\rm heat} = 0.001$ for
long dashed line (see eq. [\protect{\ref{Jpreheat}}]). The
cosmological parameters for these two latter models are the same as those
for the solid and dotted lines.
}
\begin{figure}
\par\centerline{%
\epsfxsize=0.5\columnwidth\epsfbox{\figdir/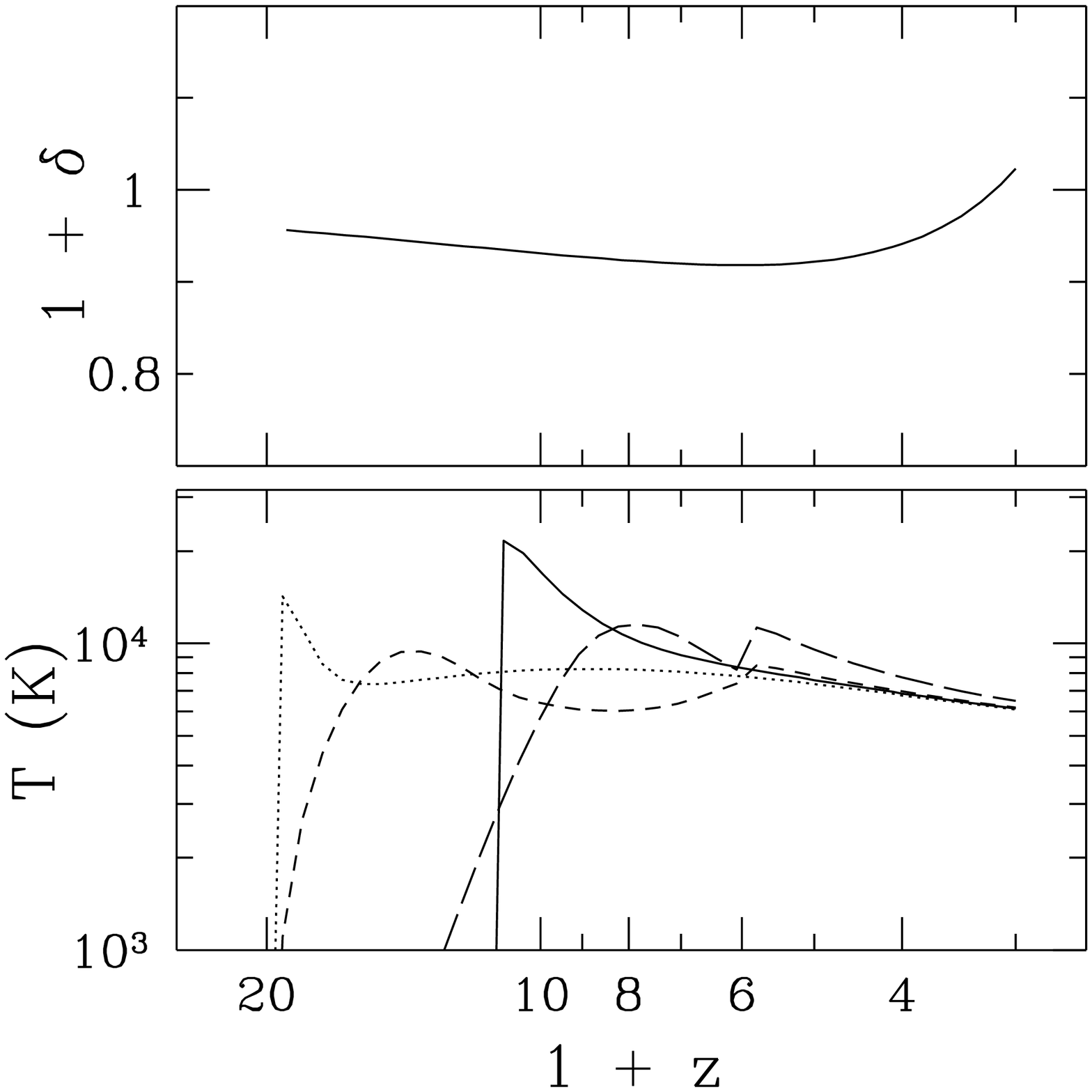}}%
\caption{\capIab}
\label{asymptote}
\end{figure}

We have tested several other reionization models where all the parameters 
in equation (\ref{Jpreheat}) are varied systematically. The same 
conclusion holds for all of them, namely that a sudden reionization model (no
preceding reheating) can always be found to fit the
temperature-density relation of any reionization model with reheating, with
the turn-on redshift of the former chosen somewhere between the epochs of
reheating and reionization of the latter. 

\section{Discussion}
\label{discuss}

Let us summarize what we have learned so far.

We find that the temperature-density relation of the low density intergalactic
medium, while having an intrinsic scatter, is well described by a mean
equation of state of the form $T = T_0 (1+\delta)^{\gamma-1}$.
It is shown that any reionization model with gradual radiation-turn-on produces
equation of state that can be well approximated by one 
produced by a suitably chosen sudden turn-on model (\S~\ref{preheat}). 

In general, the mean temperature $T_0$ at $\delta = 0$ at a given redshift
$z \sim 2-4$ decreases as the epoch of reionization is pushed earlier. It is
only weakly dependent on the amplitude of the ionizing flux $J_{\HI}$ at $z \sim
2-4$ assuming reionization takes place at $z \approxgt 5$. The spectrum of
$J_\nu$ can introduce about $20 \%$ change to $T_0$. Its main effect is
on the abundances of helium species, affecting the amount of photoionization
heating and recombination cooling. Cosmological parameters change $T_0$ in a
simple way: for sufficiently early reionization, $T_0 \propto ({\Omega_b} h /
\sqrt{\Omeganow})^{1/1.7}$.  All the dependence mentioned above are expressed in 
a quantitative way in equation (\ref{T0modified}), with about $20 \%$ accuracy
depending on the spectrum of $J_\nu$ and cosmology.

The slope of the equation of state, $\gamma - 1$, increases with redshift and
with earlier reionization epoch (eq.\ [\ref{T1}] and Fig.\ \ref{gamma}).  Its
evolution is only weakly dependent on cosmology. There
exists a maximum as to how steep $\gamma - 1$ can get, which is about $0.62$
(eq.\ [\ref{T1limit}]). Strictly speaking, $\gamma - 1$ can be as small as $0$
right at the onset of sudden reionization, but for reionization occurring
earlier than $z_{\rm reion} \sim 5$, it is greater than $0.3$ at $z = 3$;
for $z_{\rm reion}>7$ the minimum $\gamma$ is about $0.47$.

What do all these imply for the observable properties of the low column density
Ly$\alpha$ forest? Firstly, for a given density (or $1+\delta$) profile
and the Hubble constant $h$,
the column density it corresponds to is proportional to $\Omega_b^2 T^{-0.7}
J_{\HI}^{-1}$. Uncertainty in the reionization history introduces about $50 \%$
uncertainty in $T$ (see Fig.\ \ref{fig3abcd}), which in turn implies about $30
\%$ uncertainty in column density. This means an additional uncertainty in
the amplitude of the column density distribution on top of that due to
$\Omega_b$ and $J_{\HI}$, even if one is given a known cosmological model (Hui
et al.\ 1996). 

Note that for sufficiently early reionization, the dependence $T_0 \propto
{\Omega_b}^{1/1.7}$ implies that the column density is scaled by 
$\Omega_b^{2-0.7/1.7} = \Omega_b^{1.6}$, which is roughly consistent with the
result of Croft et al.\ (1996),
 who found $\Omega_b^{1.7}$ scaling using 
hydrodynamic simulations. 

The good news, on the other hand, is that the slope of the column density
distribution is relatively independent of reionization history.
The slope of the column density distribution
(number of absorption lines per unit column density per unit redshift $\propto
N_{\HI}^{-\beta}$ where $N_{\HI}$ is the column density) is approximately
given by:
\begin{equation}
\beta = 1 + {1\over {1.68 - 0.7(\gamma -1)}} [0.96 - 2 (\sigma_0 - 1) ] \,
\label{beta}
\end{equation}
where $\sigma_0$ is the rms mass fluctuation at scales close to the Jeans
length. It is close to $1$ but its precise value is determined by the
cosmological model (Hui et al.\ 1996). 
Taking the range of $\gamma - 1$ from 
$0.3$ to $0.62$ we have derived before and $\sigma_0 \sim 1$, the uncertainty
in $\beta$ introduced by unknown reionization history is of the order of $5
\%$.  

This is a very important conclusion because the slope of column density
distribution then becomes a powerful discriminant of cosmological models
through 
$\sigma_0$. It is relatively free of uncertainties due to reionization history,
or the values of $\Omega_b$ and $J_{\HI}$, unlike the amplitude of the column
density distribution. The reader is referred to Hui et al.\ (1996) for a more
detailed discussion of the predictions of various cosmological models for
$\beta$. 

The temperature-density relation also has significant implications for the
$b$ parameter distribution. There exists a rather sharp lower
cut-off of about $b = 20\dim{km}\dim{s}^{-1}$ according to high resolution
data of the low column density Ly$\alpha$ observation (Hu et al.\ 1996).
If one naively assumes the observed $b$ is completely a result of thermal
broadening $b = \sqrt{2 kT/m_p}$, one obtains a lower cut-off temperature of 
about $24000\dim{K}$. Judging from Fig.\ \ref{fig3abcd}, it is clear that
at $z \sim 3$, which is the relevant redshift for these observations, unless
the universe reionizes quite a bit later than $z = 5$, all the underdense
regions (and including some overdense regions too,
 depending on the reionization
history) have lower temperature that this cut-off. This is true even if one
allows for a lot of helium photoionization heating or very little helium
recombination cooling by varying the spectrum (see Fig.\ \ref{spectrum_fig8}).
There are a few possible resolutions: first, it
is possible that the low column density objects that we observe (column
density lower than about $10^{15}\dim{cm}^2$) actually correspond to
higher densities ($\delta$) , but that would imply a very high $J_\HI$ (higher
than about $2.0$) 
which is difficult to reconcile with observations (see Hui et al.\ 1996); 
second, it could be that the universe is
not photo-reionized after all and some other mechanism is at work;
third,
the observed $b$ parameter for these low column density systems is not the
result of thermal broadening alone. We consider the last possibility the most
likely one. As we have argued in an earlier paper (Hui et al.\ 1996) 
and was pointed
out before by numerous authors (see for instance Cen et al.\ 1994;
Hernquist et al.\ 1996; Miralda-Escud\'{e} et al.\
1996; Zhang et al.\ 1995),
the observed absorption profiles do not necessarily
conform to the classic Voigt profile but can instead reflect more the
underlying mass density profile and peculiar velocity structure. For instance,
a structure collapsing perpendicular to the line of sight (thus having density
enhancement to induce enhanced absorption) can have an expanding velocity
component along the line of sight, thus creating a profile broader than what
naive thermal broadening would predict. 

Finally, let us speculate on how a fluctuating ionizing field might change the
results of our paper. From Fig. \ref{spectrum_fig8}, it is clear that a fluctuating
field with inhomogeneous spectrum can certainly introduce a lot of scatter to the
temperature-density relation, by giving different photoionization heating and
recombination cooling rates even for fluid elements with the same density
evolution. Moreover, the ionizing field can also turn on at 
different times for different fluid elements because of radiative transfer
effects (Zuo 1992a,b). This would introduce additional scatter because as we
have seen, the earlier the reionization epoch, the lower the temperature at a
given redshift. 

However, we also learn that the temperature-density relation approaches an
asymptotic limit if one waits long enough.
For sufficiently early reionization, 
neither the amplitude of $J_\HI$ nor the reionization history (eg. at what
different times radiation turns on at different elements) affects the
temperature very much, assuming $J_\HI$ is large enough to
keep the universe reionized (see \S~\ref{spectrum}). Therefore, provided the
amplitude of 
$J_\nu$ at the helium ionizing frequencies does not vary significantly with
position at later times $z\sim 2 - 4$, the equation of state we calculate for
the early reionization models should be valid, even if the amplitude of $J_\HI$
fluctuates at $z\sim 2 - 4$. This certainly warrants further research, which we
will leave for future work. 

We are grateful to Prof.\ Martin Rees for valuable comments and suggestions
that considerably improved the manuscript.
This work was supported in part by the DOE and by the NASA (NAGW-2381) at
Fermilab and by the
UC Berkeley grant 1-443839-07427.
Simulations were performed on the NCSA Power Challenge
Array under the grant AST-960015N.
N.\ G.\ is grateful to Joshua Frieman and the Theoretical
Astrophysics group at Fermilab for hospitality, where part of this work is
completed. 

\appendix

\section{Ionization and Recombination Rates}

In this section we compile the ionization and recombination rates used
in our calculations. We believe that this compilation is sufficiently
up-to-date to reflect all recent improvements. In some of the equations below
the following notation is used:
\[
	\lambda_j = 2{T^{\rm (TR)}_{j}\over T},
\]
where $T$ is the temperature, and $T^{\rm (TR)}_{j}$ are ionization thresholds
for species $j=\HI,\GI,\GII$ expressed in temperature units,
\[
\begin{array}{lll}
	T^{\rm (TR)}_{\HI} & = & 157807\dim{K}, \\
	T^{\rm (TR)}_{\GI} & = & 285335\dim{K}, \\
	T^{\rm (TR)}_{\GII} & = & 631515\dim{K}. \\
\end{array}
\]
The symbol $k_B$ denotes the Boltzmann constant.

For chemical evolution, we use the symbols $RI$, $CI$ and $DI$ to denote
hydrogen and helium
recombination rates, collisional ionization rates and dielectronic
recombination rates.

For thermal evolution, we use the symbols $RC$, $CC$, $DC$ and $EC$ to denote
hydrogen and helium
recombination cooling, collisional ionization cooling, dielectronic
recombination cooling and line excitation cooling rates.

Superscripts and subscripts are added to the above symbols to distinguish between
different processes within each class.

Molecular hydrogen and metal cooling are not expected
to be important for the low density intergalactic medium that we are
interested in (see Songaila \& Cowie 1996 for a discussion of metals
detected in the Ly$\alpha$ forest). 

Compton cooling rates and the photoionization cross-sections are given at the
end. 

\begin{description}

\item[case A $\HII$ recombination coefficient:] our fit to the data
from Ferland et al.\ (1992); accurate to 2\% from $3\dim{K}$ to
$10^9\dim{K}$:
\[
	RI^A_{\HII} = 1.269\times10^{-13}\dim{cm}^3\dim{sec}^{-1}
	{\lambda_{\HI}^{1.503}\over
         \left(1.0+(\lambda_{\HI}/0.522)^{0.470}\right)^{1.923}}
\]
\item[case A $\HII$ recombination cooling rate:] our fit to the data
from Ferland et al.\ (1992); accurate to 2\% from $3\dim{K}$ to
$10^9\dim{K}$:
\[
	RC^A_{\HII} = 1.778\times10^{-29}\dim{erg}\dim{cm}^3\dim{sec}^{-1}
	\dim{K}^{-1}
	T
	{\lambda_{\HI}^{1.965}\over
         \left(1.0+(\lambda_{\HI}/0.541)^{0.502}\right)^{2.697}}
\]
\item[case A $\GII$ recombination coefficient:] from
from Burgess \& Seaton (1960); accurate to $\sim$10\% from $5\times10^3
\dim{K}$ to $5\times10^5\dim{K}$:
\[
	RI^A_{\GII} = 3.0\times10^{-14}\dim{cm}^3\dim{sec}^{-1}
	\lambda_{\GI}^{0.654}
\]
\item[case A $\GII$ recombination cooling rate:] from
from Burgess \& Seaton (1960); accurate to $\sim$10\% from $5\times10^3
\dim{K}$ to $5\times10^5\dim{K}$:
\[
	RC^A_{\GII} = k_BT\,RI^A_{\GII}	
\]
\item[case A $\GIII$ recombination coefficient:] our fit to the data
from Ferland et al.\ (1992); accurate to 2\% from $1\dim{K}$ to
$10^9\dim{K}$:
\[
	RI^A_{\GIII} = 2.0\times1.269\times10^{-13}\dim{cm}^3\dim{sec}^{-1}
	{\lambda_{\GII}^{1.503}\over
         \left(1.0+(\lambda_{\GII}/0.522)^{0.470}\right)^{1.923}}
\]
\item[case A $\GIII$ recombination coefficient:] our fit to the data
from Ferland et al.\ (1992); accurate to 2\% from $1\dim{K}$ to
$10^9\dim{K}$:
\[
	RC^A_{\GIII} = 8\times1.778\times10^{-29}\dim{erg}\dim{cm}^3\
	dim{sec}^{-1}\dim{K}^{-1}
	T
	{\lambda_{\GII}^{1.965}\over
         \left(1.0+(\lambda_{\GII}/0.541)^{0.502}\right)^{2.697}}
\]
\item[case B $\HII$ recombination coefficient:] our fit to the data
from Ferland et al.\ (1992); accurate to 0.7\% from $1\dim{K}$ to
$10^9\dim{K}$:
\[
	RI^B_{\HII} = 2.753\times10^{-14}\dim{cm}^3\dim{sec}^{-1}
	{\lambda_{\HI}^{1.500}\over
         \left(1.0+(\lambda_{\HI}/2.740)^{0.407}\right)^{2.242}}
\]
\item[case B $\HII$ recombination cooling rate:] our fit to the data
from Ferland et al.\ (1992); accurate to 2\% from $1\dim{K}$ to
$10^9\dim{K}$:
\[
	RC^B_{\HII} = 3.435\times10^{-30}\dim{erg}\dim{cm}^3\dim{sec}^{-1}
	\dim{K}^{-1}
	T
	{\lambda_{\HI}^{1.970}\over
         \left(1.0+(\lambda_{\HI}/2.250)^{0.376}\right)^{3.720}}
\]
\item[case B $\GII$ recombination coefficient:] from
from Burgess \& Seaton (1960); accurate to $\sim$10\% from $5\times10^3
\dim{K}$ to $5\times10^5\dim{K}$:
\[
	RI^B_{\GII} = 1.26\times10^{-14}\dim{cm}^3\dim{sec}^{-1}
	\lambda_{\GI}^{0.750}
\]
\item[case B $\GII$ recombination cooling rate:] from
from Burgess \& Seaton (1960); accurate to $\sim$10\% from $5\times10^3
\dim{K}$ to $5\times10^5\dim{K}$:
\[
	RC^B_{\GII} = k_BT\,RI^B_{\GII}	
\]
\item[case B $\GIII$ recombination coefficient:] our fit to the data
from Ferland et al.\ (1992); accurate to 2\% from $3\dim{K}$ to
$10^9\dim{K}$:
\[
	RI^B_{\GIII} = 2.0\times2.753\times10^{-14}\dim{cm}^3\dim{sec}^{-1}
	{\lambda_{\GII}^{1.500}\over
         \left(1.0+(\lambda_{\GII}/2.740)^{0.407}\right)^{2.242}}
\]
\item[case B $\GIII$ recombination cooling rate:] our fit to the data
from Ferland et al.\ (1992); accurate to 2\% from $3\dim{K}$ to
$10^9\dim{K}$:
\[
	RC^B_{\GIII} = 8\times3.435\times10^{-30}\dim{erg}\dim{cm}^3
	\dim{sec}^{-1}\dim{K}^{-1}
	T
	{\lambda_{\GII}^{1.970}\over
         \left(1.0+(\lambda_{\GII}/2.250)^{0.376}\right)^{3.720}}
\]
\item[$\HI$ collisional ionization coefficient:] our fit to the data
from Lotz (1967); accurate to 3\% from $10^4\dim{K}$ to $10^9\dim{K}$:
\[
	CI_{\HI} = 21.11\dim{cm}^3\dim{sec}^{-1}\dim{K}^{3/2}
	T^{-3/2}e^{\displaystyle -\lambda_{\HI}/2}
	{\lambda_{\HI}^{-1.089}\over
	\left(1+(\lambda_{\HI}/0.354)^{0.874}\right)^{1.101}}
\]
\item[$\HI$ collisional ionization cooling rate:] derived from
the collisional ionization coefficient:
\[
        CC_{\HI} = k_B T^{\rm (TR)}_{\HI}\, CI_{\HI}
\]
\item[$\GI$ collisional ionization coefficient:] our fit to the data
from Lotz (1967); accurate to 3\% from $10^4\dim{K}$ to $10^9\dim{K}$:
\[
	CI_{\GI} = 32.38\dim{cm}^3\dim{sec}^{-1}\dim{K}^{3/2}
	T^{-3/2}e^{\displaystyle -\lambda_{\GI}/2}
	{\lambda_{\GI}^{-1.146}\over
	\left(1+(\lambda_{\GI}/0.416)^{0.987}\right)^{1.056}}
\]
\item[$\GI$ collisional ionization cooling rate:] derived from
the collisional ionization coefficient:
\[
        CC_{\GI} = k_B T^{\rm (TR)}_{\GI}\, CI_{\GI}
\]
\item[$\GII$ collisional ionization coefficient:] our fit to the data
from Lotz (1967); accurate to 3\% from $10^4\dim{K}$ to $10^9\dim{K}$:
\[
	CI_{\GII} = 19.95\dim{cm}^3\dim{sec}^{-1}\dim{K}^{3/2}
	T^{-3/2}e^{\displaystyle -\lambda_{\GII}/2}
	{\lambda_{\GII}^{-1.089}\over
	\left(1+(\lambda_{\GII}/0.553)^{0.735}\right)^{1.275}}
\]
\item[$\GII$ collisional ionization cooling rate:] derived from
the collisional ionization coefficient:
\[
        CC_{\GII} = k_B T^{\rm (TR)}_{\GII}\, CI_{\GII}
\]
\item[$\GII$ dielectronic recombination coefficient:] from Aldrovandi
\& Pequignot (1973); accurate to $\sim$5\% from $3\times10^4\dim{K}$ 
to $10^6\dim{K}$:
\[
	DI_{\GII} = 1.90\times10^{-3}\dim{cm}^3\dim{sec}^{-1}\dim{K}^{3/2}
	T^{-3/2}e^{\displaystyle -0.75\lambda_{\GIII}/2}
	\left(1+0.3e^{\displaystyle -0.15\lambda_{\GIII}/2}\right)
\]
\item[$\GII$ dielectronic recombination cooling rate:] derived from
the dielectronic recombination coefficient:
\[
	DC_{\GII} = 0.75 k_B T^{\rm (TR)}_{\GIII}\, DI_{\GIII}
\]
\item[$\HI$ line excitation cooling rate:] from Black (1981) with the
correction from Cen (1992); 
accurate to $\sim$10\% from $5\times10^3\dim{K}$ to $5\times10^5\dim{K}$: 
\[
	EC_{\HI} = 7.5\times10^{-19}\dim{erg}\dim{cm}^3\dim{sec}^{-1}
	e^{\displaystyle -0.75\lambda_{\HI}/2}
	{1\over1+(T/10^5\dim{K})^{1/2}}
\]
\item[$\GII$ line excitation cooling rate:] from Black (1981)with the
correction from Cen (1992); 
accurate to $\sim$10\% from $5\times10^3\dim{K}$ to $5\times10^5\dim{K}$: 
\[
	EC_{\GII} = 5.54\times10^{-17}\dim{erg}\dim{cm}^3\dim{sec}^{-1}
	\left(1\dim{K}\over T\right)^{0.397}
	e^{\displaystyle -0.75\lambda_{\GII}/2}
	{1\over1+(T/10^5\dim{K})^{1/2}}
\]
\item[Compton heating/cooling term] in the equation (\ref{T}) is given
by the following expression (Peebles 1993), which is exact in the 
nonrelativistic limit:
\[
	{dQ_{\rm Compton}\over dt} = 6.35\times10^{-41}\dim{erg}\dim{cm}^{-3}
\dim{s}^{-1}\dim{K}^{-1}\Omega_bh^2\tilde{X}_e(1+z)^7\left(2.726\dim{K}
(1+z)-T\right).
\]
\end{description}

\tablefit

Photoionization cross-sections are taken from Verner et al.\ (1996)
in the following form:
\[
	\sigma(E) = \sigma_0\left[(x-1)^2+y_w^2\right]
	{y^{0.5P-5.5}\over\left(1+\sqrt{y/y_a}\right)^P},
\]
where $x=E/E_0-y_0$ and $y=\sqrt{x^2+y_1^2}$. Quantities $\sigma_0$,
$E_0$, $y_w$, $P$, $y_a$, $y_0$, and $y_1$ are fitting parameters
and are given in the Table \ref{tabfit}. They are accurate to within
10\% from the respective ionization thresholds to $5\dim{keV}$.

\end{document}